\documentclass{jpsj3}

\def\runauthor{Takanobu {\sc Jujo}}

\title{%
Relaxation Dynamics of Photoinduced Changes in 
the Superfluid Weight of High-$T_{\rm c}$ Superconductors
}

\author{%
Takanobu {\sc Jujo}
\thanks{E-mail address: jujo@ms.aist-nara.ac.jp}
}

\inst{%
Graduate School of Materials Science, Nara Institute of Science 
and Technology, Ikoma, Nara 630-0101, Japan 
}

\recdate{\today}

\abst{%
In the transient state of $d$-wave superconductors, 
we investigate the temporal variation of 
photoinduced changes in the superfluid weight. 
We derive the formula that relates the 
nonlinear response function to the nonequilibrium 
distribution function. 
The latter quantity is obtained by 
solving the kinetic equation with the electron-electron 
and the electron-phonon interaction included.  
By numerical calculations, a nonexponential decay 
is found at low temperatures 
in contrast to the usual exponential decay at 
high temperatures. 
The nonexponential decay originates 
from the nonmonotonous temporal variation 
of the nonequilibrium distribution function 
at low energies. 
The main physical process that causes this behavior 
is not the recombination of quasiparticles
as previous phenomenological studies suggested, but 
the absorption of phonons.  
}

\kword{%
nonequilibrium superconductivity, 
transient state, pump-probe, superfluid weight, 
vertex correction, unconventional superconductors
}

\begin{document}

\setlength{\textwidth}{504pt}
\setlength{\columnsep}{14pt}
\hoffset-23.5pt

\maketitle

\section{Introduction}

There have been many optical 
spectroscopic experiments on 
nonequilibrium superconductors recently, 
and they have usually been performed by pump-probe techniques. 
~\cite{averitt,segre,dvorsek,
gedik04,kaindl,perfetti,kusar,giannetti}
As for the interpretation of experimental results, 
the photoinduced change in reflectivity 
is often assumed to be directly proportional 
to the excited nonequilibrium electron density.
~\cite{kabanov,averitt,kusar}
Then the temporal evolution of the photoinduced 
quasiparticle density is analyzed with the use of 
the Rothwarf-Taylor (RT) equation,~\cite{gedik04,kaindl,giannetti}
which is first presented in ref. 10. 
This type of phenomenological equation 
is claimed to be theoretically derived 
in refs. 11 and 12 
with the use of the kinetic equation for 
the electron distribution function. 

The temporal evolution of the superfluid weight 
is observed in the experiment with the use of an optical pump 
and a THz probe,~\cite{kaindl} 
and it is found that 
the photoinduced change in the superfluid weight decays 
nonexponentially. 
The RT equation has been considered to 
be suitable for describing nonexponential relaxation 
as the bimolecular relaxation, 
because 
this equation includes the quadratic term of the nonequilibrium 
quasiparticle density 
as the recombination of the quasiparticles. 
However, the quadratic term is not allowed 
in the perturbation expansion of the external field 
for the reason that it does not satisfy 
the consistency of the equation 
with respect to the order of the pump intensity. 
Therefore, a question arises about the origin of 
the nonexponential relaxation 
in the case that the intensity of the pump beam is low. 

In this study with regard to this problem, 
we reexamine two points that have been usually 
assumed in previous works. 
One is about the relation between the photoinduced 
change in optical conductivity and the nonequilibrium 
quasiparticle density. 
This relation is not self-evident, but 
it has not been investigated so far, 
in addition to the expression of the former quantity 
itself. 
The other is about the temporal evolution of the 
nonequilibrium quasiparticle density. 
The phenomenological description given by 
the RT equation~\cite{rothwarf,kabanov05} 
is based on averaged quasiparticle density. 
This is also open to doubt for reason that 
the nonequilibrium quasiparticle density 
should have a dependence on energy. 
The kinetic equation is not simply 
averaged over energy because of the existence of 
the interaction effect as 
the kinetic equation in the normal state~\cite{ahn,groeneveld,kabanov08}
indicates. 

To investigate these problems, 
we microscopically 
calculate the response function 
of the transient state. 
Our calculation shows that the vertex correction 
term is predominant in 
photoinduced changes in the superfluid weight, 
which is same as that in the case of 
the steady state in a previous study.~\cite{jujo10} 
The nonequilibrium distribution function 
is related to the response function 
only through the interaction effect (the vertex correction term). 
The kinetic equation for this function is solved 
with the use of the electron-electron and electron-phonon 
interactions as the collision integral. 
The numerical calculation shows that, at low temperatures, 
the photoinduced change in the transient reflectivity 
is not directly proportional to 
the number of photoexcited quasiparticles, 
and the nonexponential decay does not 
originate from the recombination term, but 
from the absorption of phonons that 
enhances the nonequilibrium electron distribution. 
The effect of nonequilibrium phonons is 
also considered, and this makes the relaxation dynamics 
slow as a result of the interaction effect. 

\section{Response Function in the Transient States}

The current under the pump ($A^{\rm pu}$) 
and probe ($A^{\rm pr}$) beam is written as 
\[
J^{(3)}(t)=-
\int\frac{{\rm d}\omega}{2\pi}\int\frac{{\rm d}\Omega}{2\pi}
K^{(3)}(-\omega,-\Omega+\omega/2,\Omega+\omega/2)
A^{\rm pu}_{-\Omega+\omega/2}A^{\rm pu}_{\Omega+\omega/2}
A^{\rm pr}_{-\omega}. 
\]
Here, 
$A^{\rm pu}_{\Omega}=A'_{\rm pu}
[{\rm e}^{-\Delta T^2(\Omega-\Omega_0)^2/4}
+{\rm e}^{-\Delta T^2(\Omega+\Omega_0)^2/4}]$ and 
$A^{\rm pr}_{\omega}={\rm e}^{{\rm i}\omega t}A'_{\rm pr}
{\rm e}^{-\Delta t^2\omega^2/4}$ 
in the case of  the Gaussian pulse; 
$A^{\rm pu}_{\tau}=A_{\rm pu}
\cos(\Omega_0\tau){\rm e}^{-(\tau/\Delta T)^2}$ and 
$A^{\rm pr}_{\tau}=A_{\rm pr}
\cos(\omega_0\tau){\rm e}^{-(\tau/\Delta t)^2}$
($A'_{\rm pu}=A_{\rm pu}\sqrt{\pi}\Delta T/2$ and 
$A'_{\rm pr}=A_{\rm pr}\sqrt{\pi}\Delta t$). 
Then 
$A^{\rm pu}_{\Omega}=\int{\rm d}\tau
{\rm e}^{{\rm i}\Omega\tau}A^{\rm pu}_{\tau}$ and 
$A^{\rm pu}_{\omega}=\int{\rm d}\tau
{\rm e}^{{\rm i}\omega\tau}A^{\rm pu}_{\tau-t}$ in which 
$t$ is the time delay of the probe pulse 
after the pump excitation. 
We put $\omega_0=0$ to probe the 
the superfluid weight. 
The expression of $K^{(3)}$ is given in ref. 18, 
but the rewritten form is presented below, which clarifies 
its relation to the nonequilibrium distribution function 
and its temporal evolution. 
The integration of $K^{(3)}(-\omega,-\Omega+\omega/2,\Omega+\omega/2)$ 
by $\omega$ and $\Omega$ will lead to 
the temporal variation of the superfluid weight, if possible. 
However it is difficult to perform this integration 
numerically for the reason that 
the calculation at a very small 
energy scale is required to determine 
the variation for a long time. 
Then we transform the expression of $K^{(3)}$ 
into a feasible one, as shown below. 

We make the following assumptions 
to calculate the nonlinear response function. 
The order parameter takes a constant value and 
is not affected by the external field. 
The origin of the pairing interaction is not specified 
(the existence of the superconducting gap is assumed at the outset). 
The momentum dependence of the vertex correction is 
weak, which leads to the omission of $\Sigma^{(1)}$ and 
$\Sigma^{(3)}$ ($\Sigma^{(n)}$ is the self-energy 
that includes the external fields 
of the $n$-th order). 
We take the local approximation as discussed in 
ref. 17. 

Then the nonlinear response function 
is written as 
$K^{(3)}=N_0\int_{\rm FS}\int{\rm d}\epsilon 
v_k^4g^{(3)}_{\epsilon,\epsilon}$. 
Here, $v_k$ is the quasiparticle velocity, 
the summation $\sum_{k}$ is transformed into 
the integration $N_0\int_{\rm FS}\int{\rm d}\xi$ 
($N_0$ is the density of states at the Fermi surface), and 
$g^{(3)}_{\epsilon,\epsilon}=
\tanh\tfrac{\epsilon}{2T}
(g^{R(3)}_{\epsilon,\epsilon}-g^{A(3)}_{\epsilon,\epsilon})
+g^{(a)(3)}_{\epsilon,\epsilon}$.~\cite{note1} 
$g=\int\tfrac{{\rm d}\xi}{\pi{\rm i}}G$
(the integration of Green's function $G$
by the electron dispersion $\xi$), and 
$R$, $A$, and $(a)$ indicate the retarded, advanced, and 
anomalous parts of Green's function, respectively, 
as introduced in ref. 20, 
in which the microscopic formulation 
for the nonequilibrium superconductors 
using Green's function is given. 
$(3)$ indicates the third-order of the external fields, and 
we represent $\tilde{g}$ ($g$) as 
Green's function, which includes (does not include) 
the external fields $H_{\omega}$. 
The predominant term $g^{(a)(3)}_{\epsilon,\epsilon}$ 
is determined by 
$\tilde{g}^{(a)(3)}_{\epsilon,\epsilon}=
\int\frac{{\rm d}\omega}{2\pi}\int\frac{{\rm d}\Omega}{2\pi}
g^{(a)(3)}_{\epsilon,\epsilon}
H_{-\Omega+\omega/2}H_{\Omega+\omega/2}H_{-\omega}$, 
which satisfies the following equation 
with reference to \'Eliashberg's formulation.~\cite{eliashberg} 
\[
\begin{split}
(\epsilon^R-\epsilon^A)\tilde{g}^{(a)(3)}_{\epsilon,\epsilon}
=&
\int\frac{{\rm d}\omega}{2\pi}
[H_{-\omega}\tilde{g}^{(a)(2)}_{\epsilon,\epsilon-\omega}
-H_{\omega}\tilde{g}^{(a)(2)}_{\epsilon-\omega,\epsilon}] 
+
\int\frac{{\rm d}\omega}{2\pi}
[H_{-\omega}(T_{\epsilon-\omega}-T_{\epsilon})
\tilde{g}^{R(2)}_{\epsilon,\epsilon-\omega}
-H_{\omega}(T_{\epsilon}-T_{\epsilon-\omega})
\tilde{g}^{A(2)}_{\epsilon-\omega,\epsilon}] \\
&+
\int\frac{{\rm d}\omega}{2\pi}
[
\Sigma^{R(2)}_{\epsilon,\epsilon-\omega}
\tilde{g}^{(a)(1)}_{\epsilon-\omega,\epsilon}
-\tilde{g}^{(a)(1)}_{\epsilon,\epsilon-\omega}
\Sigma^{A(2)}_{\epsilon-\omega,\epsilon}
+\Sigma^{(a)(2)}_{\epsilon,\epsilon-\omega}
\tilde{g}^{A(1)}_{\epsilon-\omega,\epsilon}
-\tilde{g}^{R(1)}_{\epsilon,\epsilon-\omega}
\Sigma^{(a)(2)}_{\epsilon-\omega,\epsilon}
]. 
\end{split}
\]
Here, $T_{\epsilon}=\tanh\frac{\epsilon}{2T}$, 
$\epsilon^{R(A)}=\epsilon-\Sigma^{R(A)}_{k,\epsilon}$, 
and $H_{\omega}=v_kA_{\omega}$ with 
the external field $A_{\omega}$. 
The functions $\tilde{g}$ in this equation 
satisfy similar equations. 
By taking only the predominant terms into account, 
$\tilde{g}^{(a)(3)}_{\epsilon,\epsilon}$ is written as 
\begin{equation}
\begin{split}
\tilde{g}^{(a)(3)}_{\epsilon,\epsilon}
\simeq &
\frac{1}{\epsilon^R-\epsilon^A}
\int\frac{{\rm d}\omega}{2\pi}
\int\frac{{\rm d}\Omega}{2\pi}
[
R_{\omega}(\Omega)
g^{R(1)}_{\epsilon,\epsilon+\omega}
+
R^*_{-\omega}(\Omega)
g^{A(1)}_{\epsilon-\omega,\epsilon}
]
H_{-\Omega+\omega/2}H_{\Omega+\omega/2}H_{-\omega} \\
&+
\frac{1}{\epsilon^R-\epsilon^A}
\int\frac{{\rm d}\omega}{2\pi}
[\Sigma^{(a)(2)}_{\epsilon,\epsilon+\omega}g^{A(1)}_{\epsilon+\omega,\epsilon}
-g^{R(1)}_{\epsilon,\epsilon+\omega}\Sigma^{(a)(2)}_{\epsilon+\omega,\epsilon}
]H_{-\omega}. 
\label{eq:ga3}
\end{split}
\end{equation}
Here, 
\[
R_{\omega}(\Omega)=
\frac{T_{\epsilon+\omega}-T_{\epsilon-\Omega+\omega/2}}
{\epsilon^R-(\epsilon-\Omega+\omega/2)^A}
+
\frac{T_{\epsilon-\Omega+\omega/2}-T_{\epsilon}}
{\epsilon^R-(\epsilon-\Omega+\omega/2)^R}
+
\frac{T_{\epsilon+\omega}-T_{\epsilon+\Omega+\omega/2}}
{\epsilon^R-(\epsilon+\Omega+\omega/2)^A}
+
\frac{T_{\epsilon+\Omega+\omega/2}-T_{\epsilon}}
{\epsilon^R-(\epsilon+\Omega+\omega/2)^R}. 
\] 
To derive the temporal variation of $K^{(3)}$, 
we consider the following integration with 
use of some function $f_{\omega}(\Omega)$; 
\begin{equation}
\int\frac{{\rm d}\omega}{2\pi}\int\frac{{\rm d}\Omega}{2\pi}
A^{\rm pu}_{-\Omega+\omega/2}A^{\rm pu}_{\Omega+\omega/2}
A^{\rm pr}_{-\omega}
f_{\omega}(\Omega)
\simeq 
\frac{A^{'2}_{\rm pu}A'_{\rm pr}
{\rm exp}[-t^2/(\Delta t^2+\Delta T^2/2)]}
{\sqrt{2}\pi\Delta T\sqrt{\Delta t^2+\Delta T^2/2}}
[f_{0}(\Omega_0)+f_{0}(-\Omega_0)]. 
\label{eq:integA}
\end{equation}
If the pulse widths $\Delta t$ and $\Delta T$ are 
several femtoseconds, 
the effective range of frequency is of the order of 
the superconducting gap ($\Delta_0$). 
Here, we consider the case that 
$\Omega_0\Delta T \gg 1$, $\Omega_0 \gg \Delta_0$, and 
the dependences of $f_{\omega}(\Omega)$ 
on $\omega$ and $\Omega$ are weak 
within $|\omega|<1/\Delta t$ and 
$|\Omega\mp\Omega_0|<1/\Delta T$, respectively. 
Firstly, we examine the temporal variation of 
the first term of eq. (\ref{eq:ga3}). 
If we take account of 
the above consideration and 
the weak dependence of this term on $\omega$, 
we can apply the integration 
eq. (\ref{eq:integA}) to this term. 
This term is negligible 
in comparison with the vertex correction term 
discussed below because of 
the smallness of $R_0(\Omega_0)$, which is 
the same as that in the case of ref. 17, 
in addition to the exponential factor. 
Next, we consider the second term of eq. (\ref{eq:ga3}) 
(the vertex correction term). 
By using the relations 
$\tilde{g}^{R(1)}_{\epsilon,\epsilon+\omega}\simeq H_{-\omega}
\partial g^R_{k,\epsilon}/\partial \epsilon$ and 
$\partial g^A_{k,\epsilon}/\partial \epsilon
=-(\partial g^R_{k,\epsilon}/\partial \epsilon)^*$, 
the vertex correction term is written as 
\[
\frac{1}{\epsilon^R-\epsilon^A}
\int\frac{{\rm d}\omega}{2\pi}
[\Sigma^{(a)(2)}_{\epsilon,\epsilon+\omega}g^{A(1)}_{\epsilon+\omega,\epsilon}
-g^{R(1)}_{\epsilon,\epsilon+\omega}\Sigma^{(a)(2)}_{\epsilon+\omega,\epsilon}
]H_{-\omega}
\simeq  
\frac{-2}{\epsilon^R-\epsilon^A}
\int\frac{{\rm d}\omega}{2\pi}
{\rm Re}\left(
\frac{\partial g^R_{k,\epsilon}}{\partial\epsilon}
\right)
\Sigma^{(a)(2)}_{k,\epsilon}(\omega)H_{-\omega}. 
\]
Here, we put 
$\Sigma^{(a)(2)}_{\epsilon+\omega/2,\epsilon-\omega/2}
=\Sigma^{(a)(2)}_{k,\epsilon}(\omega)$, 
which is a functional of 
$\tilde{g}^{(a)(2)}_{k,\epsilon}(\omega)$, and 
its functional form is determined by 
specifying the interaction and the self-energy. 
$\tilde{g}^{(a)(2)}_{k,\epsilon}(\omega)$ satisfies 
the following equation: 
\begin{equation}
\begin{split}
[(\epsilon+\omega/2)^R-(\epsilon-\omega/2)^A] 
\tilde{g}^{(a)(2)}_{k,\epsilon}(\omega)
=&
\int\frac{{\rm d}\Omega}{2\pi}
v_k^2\tilde{R}_{k,\epsilon}(\omega,\Omega)
A^{\rm pu}_{-\Omega+\omega/2}A^{\rm pu}_{\Omega+\omega/2}
-
(g^R_{k,\epsilon+\omega/2}-g^A_{k,\epsilon-\omega/2})
\Sigma^{(a)(2)}_{k,\epsilon}(\omega) \\ 
&-(f^R_{k,\epsilon+\omega/2}-f^A_{k,\epsilon-\omega/2})
\Upsilon^{(a)(2)}_{k,\epsilon}(\omega)
+(\Upsilon^R_{k,\epsilon+\omega/2}-\Upsilon^A_{k,\epsilon-\omega/2})
\tilde{f}^{(a)(2)}_{k,\epsilon}(\omega). 
\label{eq:ga2kineq}
\end{split}
\end{equation}
Here, 
$\Upsilon$ represents the anomalous (off-diagonal) 
part of the self-energy, 
$g^R_{k,\epsilon}=-2\epsilon^R/Z^R_{k,\epsilon}$ and 
$f^R_{k,\epsilon}=-2\Delta_k/Z^R_{k,\epsilon}$ 
$\left(Z^R_{k,\epsilon}=
{\rm sgn}(\epsilon)\sqrt{(\epsilon^R)^2-\Delta_k^2}\right)$, and 
$\tilde{R}_{k,\epsilon}(\omega,\Omega)
=R_{k,\epsilon}(\omega,\Omega)+R_{k,\epsilon}(\omega,-\Omega)$. 
\[
\begin{split}
R_{k,\epsilon}(\omega,\Omega)
=&
\frac{(T_{\epsilon+\omega/2}-T_{\epsilon+\Omega})
(g^R_{k,\epsilon+\omega/2}-g^A_{k,\epsilon+\Omega})}
{(\epsilon+\omega/2)^R-(\epsilon+\Omega)^A}
+
\frac{(T_{\epsilon+\Omega}-T_{\epsilon-\omega/2})
(g^R_{k,\epsilon+\omega/2}-g^R_{k,\epsilon+\Omega})}
{(\epsilon-\omega/2)^R-(\epsilon+\Omega)^R}  \\
&-
\frac{(T_{\epsilon+\omega/2}-T_{\epsilon+\Omega})
(g^A_{k,\epsilon+\Omega}-g^A_{k,\epsilon-\omega/2})}
{(\epsilon+\Omega)^A-(\epsilon-\omega/2)^A}
-
\frac{(T_{\epsilon+\Omega}-T_{\epsilon-\omega/2})
(g^R_{k,\epsilon+\Omega}-g^A_{k,\epsilon-\omega/2})}
{(\epsilon+\Omega)^R-(\epsilon-\omega/2)^A}. 
\end{split}
\]
This latter quantity is related to 
the pump-induced term and the initial nonequilibrium 
distribution function, which directly reflects 
the values of the self-energy. 
This induced term vanishes 
in the case that the self-energy is absent, 
as indicated in ref. 17. 

As shown in eq. (\ref{eq:ga2kineq}) 
the dependence of $\tilde{g}^{(a)(2)}_{k,\epsilon}(\omega)$ 
(and $\Sigma^{(a)(2)}_{k,\epsilon}(\omega)$) 
on $\omega$ is strong, and this makes the $t$-dependence 
of the vertex correction term different from that of 
the first term of eq. (\ref{eq:ga3}) and larger than that. 
By using the integration 
$\int\frac{{\rm d}\omega}{2\pi}
\tilde{g}^{(a)(2)}_{k,\epsilon}(\omega)
{\rm e}^{-{\rm i}\omega t}{\rm e}^{-\Delta t^2\omega^2/4}=
\frac{1}{\sqrt{\pi}\Delta t}\int{\rm d}\tau
\tilde{g}^{(a)(2)}_{k,\epsilon}(\tau)
{\rm e}^{-(t-\tau)^2/\Delta t^2}\simeq 
\tilde{g}^{(a)(2)}_{k,\epsilon}(t)$, 
which is applicable to 
the time range $t\gg \Delta t$, 
we can transform eq. (\ref{eq:ga2kineq}) 
into the following kinetic equation, which 
describes the long time behavior 
of $\tilde{g}^{(a)(2)}_{k,\epsilon}(t)$: 
\begin{equation}
\begin{split}
{\rm i}\frac{\partial \tilde{g}^{(a)(2)}_{k,\epsilon}(t)}{\partial t}
=&
\alpha(t)v_k^2
\tilde{R}_{k,\epsilon}(0,\Omega_0)
+(\Sigma^R_{k,\epsilon}-\Sigma^A_{k,\epsilon})
\tilde{g}^{(a)(2)}_{k,\epsilon}(t). 
-(g^R_{k,\epsilon}-g^A_{k,\epsilon})\Sigma^{(a)(2)}_{k,\epsilon}(t) \\
&+(\Upsilon^R_{k,\epsilon}-\Upsilon^A_{k,\epsilon})
\tilde{f}^{(a)(2)}_{k,\epsilon}(t). 
-(f^R_{k,\epsilon}-f^A_{k,\epsilon})\Upsilon^{(a)(2)}_{k,\epsilon}(t). 
\label{eq:tga2kineq}
\end{split}
\end{equation}
Here, 
$\alpha(t)=
\frac{A^{'2}_{\rm pu}
{\rm exp}[-t^2/(\Delta t^2+\Delta T^2/2)]}
{\sqrt{2}\pi\Delta T\sqrt{\Delta t^2+\Delta T^2/2}}$. 
To obtain $K^{(3)}$, 
the solution of this kinetic equation is 
substituted into $\Sigma^{(a)(2)}_{k,\epsilon}(t)$ 
in the following vertex correction term: 
\begin{equation}
J^{(3)}_{\rm vc}(t)=-
N_0\int_{\rm FS}\int{\rm d}\epsilon v_k^2
\frac{-2}{\epsilon^R-\epsilon^A}
{\rm Re}\left(
\frac{\partial g^R_{k,\epsilon}}{\partial\epsilon}
\right)
\Sigma^{(a)(2)}_{k,\epsilon}(t)A'_{\rm pr}. 
\label{eq:j3vc}
\end{equation}

The kinetic equation for nonequilibrium phonons 
is similarly written as 
\begin{equation}
{\rm i}
\frac{2\omega}{\omega_{\phi}^2}
\frac{\partial D^{(a)(2)}_{\omega}(t)}{\partial t}
=
(\Pi^R_{\omega}-\Pi^A_{\omega})
D^{(a)(2)}_{\omega}(t). 
-(D^R_{\omega}-D^A_{\omega})\Pi^{(a)(2)}_{\omega}(t). 
\label{eq:tda2kineq}
\end{equation}
Here, $D_{\omega}^{R,(A)}$ and 
$\Pi_{\omega}^{R,(A)}$ 
are the retarded (advanced) phonon 
Green's function and the self-energy by 
phonon-electron interaction, respectively. 
$(a)$ and $(2)$ indicate the anomalous part and 
the order of the external field, respectively, 
which is same as that in the case of electrons. 

\section{Kinetic Equation of the Nonequilibrium Distribution 
Function}

In this section, we derive the kinetic equations 
for the distribution functions of electrons and phonons. 
We put the deviation of the distribution function 
from the equilibrium state by the pump excitation 
as $\delta n_{\epsilon}(t)$ and $\delta N_{\omega_{\phi}}(t)$ 
for electrons and phonons, respectively. 
We consider a two-dimensional system and 
replace $\int_{\rm FS}$ 
by $\int\frac{{\rm d}\varphi}{2\pi}$. 
$v_k=v_{\rm F}\cos\varphi$ 
($v_{\rm F}$ is the Fermi velocity) 
and $\Delta_k=\Delta_0\cos2\varphi$ 
for $d$-wave superconductors. 

We adopt the second-order perturbation expansion 
for the electron-electron interaction 
and the one-loop approximation for the electron-phonon interaction 
as the self-energy. 
We rewrite the nonequilibrium Green function as 
$\tilde{g}^{(a)(2)}_{\varphi,\epsilon}(t)=-2\delta n_{\epsilon}(t)
(g^R_{\varphi,\epsilon}-g^A_{\varphi,\epsilon})$, 
$\tilde{f}^{(a)(2)}_{\varphi,\epsilon}(t)=-2\delta n_{\epsilon}(t)
(f^R_{\varphi,\epsilon}-f^A_{\varphi,\epsilon})$, 
and 
$D^{(a)(2)}_{\omega}(t)=
2\delta N_{\omega}(t)(D^R_{\omega}-D^A_{\omega})$ 
with the use of 
$D^R_{\omega}-D^A_{\omega}=
-\pi{\rm i}\omega
[\delta(\omega-\omega_{\phi})+\delta(\omega+\omega_{\phi})]$ 
(for example, see ref. 21). 
We consider acoustic phonons and 
put $\omega_{\phi}=v_{\rm s}k_{\rm F}|\phi|$
($v_{\rm s}$ is the sound velocity). 
Then the kinetic equation for electrons, 
eq. (\ref{eq:tga2kineq}), is written as 
\begin{equation}
\begin{split}
\frac{\partial \delta n_{\epsilon}(t)}{\partial t}
=&
\frac{1}{\bar{g}^R_{\epsilon}}
\frac{\alpha(t)}{4{\rm i}}
\int\frac{{\rm d}\varphi}{2\pi}
v_{\varphi}^2
\tilde{R}_{\varphi,\epsilon}(0,\Omega_0)
-\frac{1}{\bar{g}^R_{\epsilon}}
\frac{\pi}{2}N_0g^2\int_{-\phi_D}^{\phi_D}\frac{{\rm d}\phi}{2\pi}
\omega_{\phi}I^{\rm el-ph}_{\phi,\epsilon}
[\delta n,\delta N]  \\
&-\frac{1}{\bar{g}^R_{\epsilon}}
\frac{\pi}{2}\frac{U^2N_0^2}{v_{\rm F}k_{\rm F}}
\iiint\frac{{\rm d}\varphi{\rm d}\varphi_1{\rm d}\varphi_2}{(2\pi)^3}
\iint{\rm d}\epsilon_1{\rm d}\epsilon_2
\tilde{\delta}_{\varphi,\varphi_1,\varphi_2}
I^{\rm el-el}_{\epsilon,\varphi;\epsilon_1,\varphi_1;\epsilon_2,\varphi_2}
[\delta n]. 
\end{split}
\label{eq:nedkineq}
\end{equation}
Here, 
$\bar{g}_{\epsilon}:=
-\int\frac{{\rm d}\varphi}{2\pi}
{\rm Re}g^R_{\varphi}(\epsilon)$ and 
$\tilde{\delta}_{\varphi,\varphi_1,\varphi_2}:=
\delta[1-\cos(\varphi-\varphi_1)+\cos(\varphi-\varphi_2)
-\cos(\varphi_1-\varphi_2)]$ is a delta function. 
$U$ is the effective short-range Coulomb repulsion 
energy, which is called the on-site Coulomb interaction 
in the Hubbard model. 
$g$ is the coefficient of the electron-phonon 
interaction, which is usually written as 
$g=\sqrt{\hbar/2M_{\rm i}v_{\rm s}}V_{\rm i}$ 
($M_{\rm i}$ is the mass of the ion and $V_{\rm i}$ 
is the renormalized ion potential). 
In the next section, 
we use $UN_0$ and $g^2N_0$ as dimensionless parameters 
that characterize the strengths of the electron-electron 
and electron-phonon interactions, respectively. 
The collision terms for the electron-electron 
and electron-phonon interactions are written as 
\[
\begin{split}
I^{\rm el-el}_{\epsilon,\varphi;\epsilon_1,\varphi_1;\epsilon_2,\varphi_2}
[\delta n]
=&
\frac{
{\rm Re}g^R_{\varphi,\epsilon}{\rm Re}g^R_{\varphi_1,\epsilon_1}
{\rm Re}g^R_{\varphi_2,\epsilon_2}
{\rm Re}g^R_{\varphi-\varphi_1+\varphi_2,\epsilon-\epsilon_1+\epsilon_2}}
{\cosh\frac{\epsilon}{2T}\cosh\frac{\epsilon_1}{2T}
\cosh\frac{\epsilon_2}{2T}\cosh\frac{\epsilon-\epsilon_1+\epsilon_2}{2T}} \\
&\times
\left[
(\cosh\tfrac{\epsilon}{2T})^2\delta n_{\epsilon}(t)
-(\cosh\tfrac{\epsilon_1}{2T})^2\delta n_{\epsilon_1}(t)
+(\cosh\tfrac{\epsilon_2}{2T})^2\delta n_{\epsilon_2}(t)
-(\cosh\tfrac{\epsilon-\epsilon_1+\epsilon_2}{2T})^2
\delta n_{\epsilon-\epsilon_1+\epsilon_2}(t)
\right], 
\end{split}
\]
and 
\[
\begin{split}
I^{\rm el-ph}_{\phi,\epsilon}
[\delta n,\delta N]
=&
\left(\coth\tfrac{\omega_{\phi}}{2T}
+\tanh\tfrac{\epsilon-\omega_{\phi}}{2T}\right)
g^-_{\phi,\epsilon}\delta n_{\epsilon}(t)
+\left(\coth\tfrac{\omega_{\phi}}{2T}
-\tanh\tfrac{\epsilon+\omega_{\phi}}{2T}\right)
g^+_{\phi,\epsilon}\delta n_{\epsilon}(t) \\
&-
\left(\coth\tfrac{\omega_{\phi}}{2T}
-\tanh\tfrac{\epsilon}{2T}\right)
g^-_{\phi,\epsilon}\delta n_{\epsilon-\omega_{\phi}}(t)
-\left(\coth\tfrac{\omega_{\phi}}{2T}
+\tanh\tfrac{\epsilon}{2T}\right)
g^+_{\phi,\epsilon}\delta n_{\epsilon+\omega_{\phi}}(t) \\
&+
\left(\tanh\tfrac{\epsilon-\omega_{\phi}}{2T}
-\tanh\tfrac{\epsilon}{2T}\right)
g^-_{\phi,\epsilon}\delta N_{\phi}(t)
+\left(\tanh\tfrac{\epsilon+\omega_{\phi}}{2T}
-\tanh\tfrac{\epsilon}{2T}\right)
g^+_{\phi,\epsilon}\delta N_{\phi}(t). 
\end{split}
\]
Here, 
$g^{\mp}_{\phi,\epsilon}:=
\int\frac{{\rm d}\varphi}{2\pi}
\left[
{\rm Re}g^R_{\varphi-\phi}(\epsilon\mp\omega_{\phi})
{\rm Re}g^R_{\phi}(\epsilon)
-{\rm Re}f^R_{\varphi-\phi}(\epsilon\mp\omega_{\phi})
{\rm Re}f^R_{\phi}(\epsilon)
\right]$. 

The kinetic equation for nonequilibrium phonons, 
eq. (\ref{eq:tda2kineq}), is written as 
\[
\frac{\partial \delta N_{\omega_{\phi}}}{\partial t}
=-\gamma_{\rm esc}\delta N_{\omega_{\phi}}
+\frac{v_{\rm s}}{v_{\rm F}}\frac{1}{2}N_0g^2
\int_0^{\infty}{\rm d}{\epsilon}
I^{\rm ph-el}_{\phi,\epsilon}
[\delta n_{\epsilon},\delta N_{\phi}]. 
\]
Here, we add a phenomenological term, 
$\gamma_{\rm esc}\delta N_{\omega_{\phi}}$
(damping by phonon escape), which 
describes the equilibration between the 
electron-phonon system and the reservoir. 
\[
\begin{split}
I^{\rm ph-el}_{\phi,\epsilon}
[\delta n_{\epsilon},\delta N_{\phi}]
=&
\left(\coth\tfrac{\omega_{\phi}}{2T}
+\tanh\tfrac{\epsilon-\omega_{\phi}}{2T}\right)
g^-_{\phi,\epsilon}\delta n_{\epsilon}(t)
-\left(\coth\tfrac{\omega_{\phi}}{2T}
-\tanh\tfrac{\epsilon+\omega_{\phi}}{2T}\right)
g^+_{\phi,\epsilon}\delta n_{\epsilon}(t) \\
&-
\left(\coth\tfrac{\omega_{\phi}}{2T}
-\tanh\tfrac{\epsilon}{2T}\right)
g^-_{\phi,\epsilon}\delta n_{\epsilon-\omega_{\phi}}(t)
+\left(\coth\tfrac{\omega_{\phi}}{2T}
+\tanh\tfrac{\epsilon}{2T}\right)
g^+_{\phi,\epsilon}\delta n_{\epsilon+\omega_{\phi}}(t) \\
&+
\left(\tanh\tfrac{\epsilon-\omega_{\phi}}{2T}
-\tanh\tfrac{\epsilon}{2T}\right)
g^-_{\phi,\epsilon}\delta N_{\phi}(t)
-\left(\tanh\tfrac{\epsilon+\omega_{\phi}}{2T}
-\tanh\tfrac{\epsilon}{2T}\right)
g^+_{\phi,\epsilon}\delta N_{\phi}(t). 
\end{split}
\]

The energy conservation is also discussed using 
the kinetic equations. 
The additional energy induced by the external field 
in the system of electrons and phonons is given by 
\[
2\sum_k
E_k\delta n_{E_k}(t)
+
\sum_q
\omega_q\delta N_{\omega_q}(t)
=
2N_0\int\frac{{\rm d}\varphi}{2\pi}\int{\rm d}\epsilon
\frac{|\epsilon|}{\sqrt{\epsilon^2-\Delta_{\varphi}^2}}
\epsilon \delta n_{\epsilon}(t)
+
\frac{N_0}{2}\int_{-\phi_D}^{\phi_D}\frac{{\rm d}\phi}{2\pi}
2\pi v_{\rm F}k_{\rm F}|\phi|
\omega_{\phi}\delta N_{\phi}(t). 
\]
With the use of the kinetic equation, it is shown 
that this quantity is equal to the energy injected 
by the external field, 
which is written as 
\[
\int_{-\infty}^{t}{\rm d}t'
N_0\int{\rm d}\epsilon\int\frac{{\rm d}\varphi}{2\pi}\frac{1}{4{\rm i}}
\alpha(t')
v_{\varphi}^2
\tilde{R}_{\varphi,\epsilon}(0,\Omega_0). 
\]
In the above discussion, we put $\gamma_{\rm esc}=0$ 
for the energy conservation. 
In the case of $\gamma_{\rm esc}\ne 0$, 
the energy of the electron-phonon system dissipates 
into the reservoir system. 

The solution of the kinetic equations is 
related to the nonlinear response function as follows. 
$\delta n_{\epsilon}(t)$ and $\delta N_{\omega}(t)$ 
obtained by solving the kinetic equations is substituted 
to $\Sigma^{(a)(2)}_{\varphi,\epsilon}(t)$ of the previous section 
as 
$\Sigma^{(a)(2)}_{\varphi,\epsilon}(t)=
\Sigma^{(a)\rm el-el}_{\varphi,\epsilon}(t)+
\Sigma^{(a)\rm el-ph}_{\varphi,\epsilon}(t)$. 
Here, 
\[
\begin{split}
\Sigma^{(a)\rm el-el}_{\varphi,\epsilon}(t)
=&-\pi{\rm i}
\frac{U^2N_0^2}{v_{\rm F}k_{\rm F}}
\iint\frac{{\rm d}\varphi_1{\rm d}\varphi_2}{(2\pi)^2}
\iint{\rm d}\epsilon_1{\rm d}\epsilon_2
\tilde{\delta}_{\varphi,\varphi_1,\varphi_2}
\frac{
{\rm Re}g^R_{\varphi_1,\epsilon_1}{\rm Re}g^R_{\varphi_2,\epsilon_2}
{\rm Re}g^R_{\varphi-\varphi_1+\varphi_2,\epsilon-\epsilon_1+\epsilon_2}}
{\cosh\tfrac{\epsilon}{2T}\cosh\tfrac{\epsilon_1}{2T}
\cosh\tfrac{\epsilon_2}{2T}
\cosh\tfrac{\epsilon-\epsilon_1+\epsilon_2}{2T}} \\
&\times
\left[
(\cosh\tfrac{\epsilon_1}{2T})^2\delta n_{\epsilon_1}(t)
-(\cosh\tfrac{\epsilon_2}{2T})^2\delta n_{\epsilon_2}(t)
+(\cosh\tfrac{\epsilon-\epsilon_1+\epsilon_2}{2T})^2
\delta n_{\epsilon-\epsilon_1+\epsilon_2}(t) 
\right], 
\end{split}
\]
and 
\[
\begin{split}
\Sigma^{(a)\rm el-ph}_{\varphi,\epsilon}(t)
=&-\pi{\rm i}
N_0g^2\int_{-\phi_D}^{\phi_D}\frac{{\rm d}\phi}{2\pi}
\omega_{\phi}\{
{\rm Re}g^R_{\varphi-\phi,\epsilon-\omega_{\phi}}
[
\left(\coth\tfrac{\omega_{\phi}}{2T}-\tanh\tfrac{\epsilon}{2T}
\right)\delta n_{\epsilon-\omega_{\phi}}(t)
-\left(\tanh\tfrac{\epsilon-\omega_{\phi}}{2T}-\tanh\tfrac{\epsilon}{2T}
\right)\delta N_{\phi}(t)] \\
&+{\rm Re}g^R_{\varphi-\phi,\epsilon+\omega_{\phi}}
[
\left(\coth\tfrac{\omega_{\phi}}{2T}+\tanh\tfrac{\epsilon}{2T}
\right)\delta n_{\epsilon+\omega_{\phi}}(t)
-\left(\tanh\tfrac{\epsilon+\omega_{\phi}}{2T}-\tanh\tfrac{\epsilon}{2T}
\right)\delta N_{\phi}(t)
] 
\}. 
\end{split}
\]
Then this $\Sigma^{(a)(2)}_{\varphi,\epsilon}(t)$ 
is substituted into eq. (\ref{eq:j3vc}), which gives 
the nonlinear response function. 

\section{Results}

The results of the numerical calculation are 
shown below. 
The superconducting gap $\Delta_0$ is taken as 
the unit of energy $\Delta_0=1.0$ 
(this leads to the superconducting transition 
temperature $T_{\rm c}=0.465$), 
and we put $\omega_{\phi}\le v_{\rm s}k_{\rm F}$. 
We fix the values of several parameters 
as follows: 
$\Omega_0=6.0$~\cite{note2}, 
$v_{\rm s}k_{\rm F}=2.0372$, 
$v_{\rm s}/v_{\rm F}=0.05$ 
(which leads to the Fermi energy 
$E_{\rm F}=v_{\rm F}k_{\rm F}/2\simeq 20$), 
$UN_0=0.2$, and $g^2N_0=0.05$ 
(this value corresponds to 
$\lambda=0.1$ as the electron-phonon coupling constant). 
We add a small $\delta=-0.01$ to 
the imaginary part of the self-energy as 
effective impurities and a finite mean free path. 

In the calculation of the kinetic equation eq. (\ref{eq:nedkineq}), 
the following approximation is used 
to make a multiple integration in the 
electron-electron collision term a feasible one: 
$\iiint\frac{{\rm d}\varphi{\rm d}\varphi_1{\rm d}\varphi_2}{(2\pi)^3}
\tilde{\delta}_{\varphi,\varphi_1,\varphi_2}
{\rm Re}g^R_{\varphi,\epsilon}{\rm Re}g^R_{\varphi_1,\epsilon_1}
{\rm Re}g^R_{\varphi_2,\epsilon_2}
{\rm Re}g^R_{\varphi-\varphi_1+\varphi_2,\epsilon-\epsilon_1+\epsilon_2}
\to
{\rm Re}\bar{g}^R_{\epsilon}{\rm Re}\bar{g}^R_{\epsilon_1}
{\rm Re}\bar{g}^R_{\epsilon_2}
{\rm Re}\bar{g}^R_{\epsilon-\epsilon_1+\epsilon_2}$. 
We perform the same approximation for 
$\Sigma^{(a){\rm el-el}}_{\varphi,\epsilon}$. 
This leads to violation of the momentum conservation, 
although the energy conservation is satisfied. 
The numerical calculation indicates that, 
by this replacement, 
the interaction effect is underestimated 
at low energies ($|\epsilon| < \Delta_0$) 
owing to averaging the angular dependence. 
However, the quantitative difference is small 
especially at $|\epsilon| > \Delta_0$. 
Then it is not considered that this approximation 
causes qualitative changes for numerical 
results. 

Below we consider two typical cases 
for a nonequilibrium phonon system. 
The numerical calculation of 
$K^{(3)}$ shows 
that calculations with large values of 
the damping $\gamma_{\rm esc}$ 
lead to results 
similar to those in the case of $\delta N=0$ 
(phonons in thermal equilibrium). 
On the other hand, 
the results with small values of $\gamma_{\rm esc}$ 
are approximated by those for $\gamma_{\rm esc}=0$. 
Therefore, 
we show the calculated results 
only for $\delta N=0$ and $\delta N\ne 0$ with $\gamma_{\rm esc}=0.0$ 
as characteristic cases of the phonon system. 

To solve the kinetic equations, 
we consider the square pulse 
and put $\alpha(t)=1$ for $0<t<0.25$ and 
$\alpha(t)=0$ otherwise. 
(The long time behavior is not affected if we use 
a Gaussian pulse.) 
Then we redefine $K^{(3)}$ as 
$K^{(3)}=-J_{\rm vc}^{(3)}(t)/A'_{\rm pr}$
from eq. (\ref{eq:j3vc}) with the use of this $\alpha(t)$, 
and this is the photoinduced change in the superfluid weight. 
(We can put $J^{(3)}(t)\simeq J^{(3)}_{\rm vc}(t)$ 
as noted above.) 
If we use only $\Sigma^{(a){\rm el-el}}$ ($\Sigma^{(a){\rm el-ph}}$)
as $\Sigma^{(a)(2)}$ in this equation, 
we write this quantity as 
$K^{(3){\rm el-el}}$ ($K^{(3){\rm el-ph}}$) below. 
 
\begin{figure}
\includegraphics[width=8.5cm]{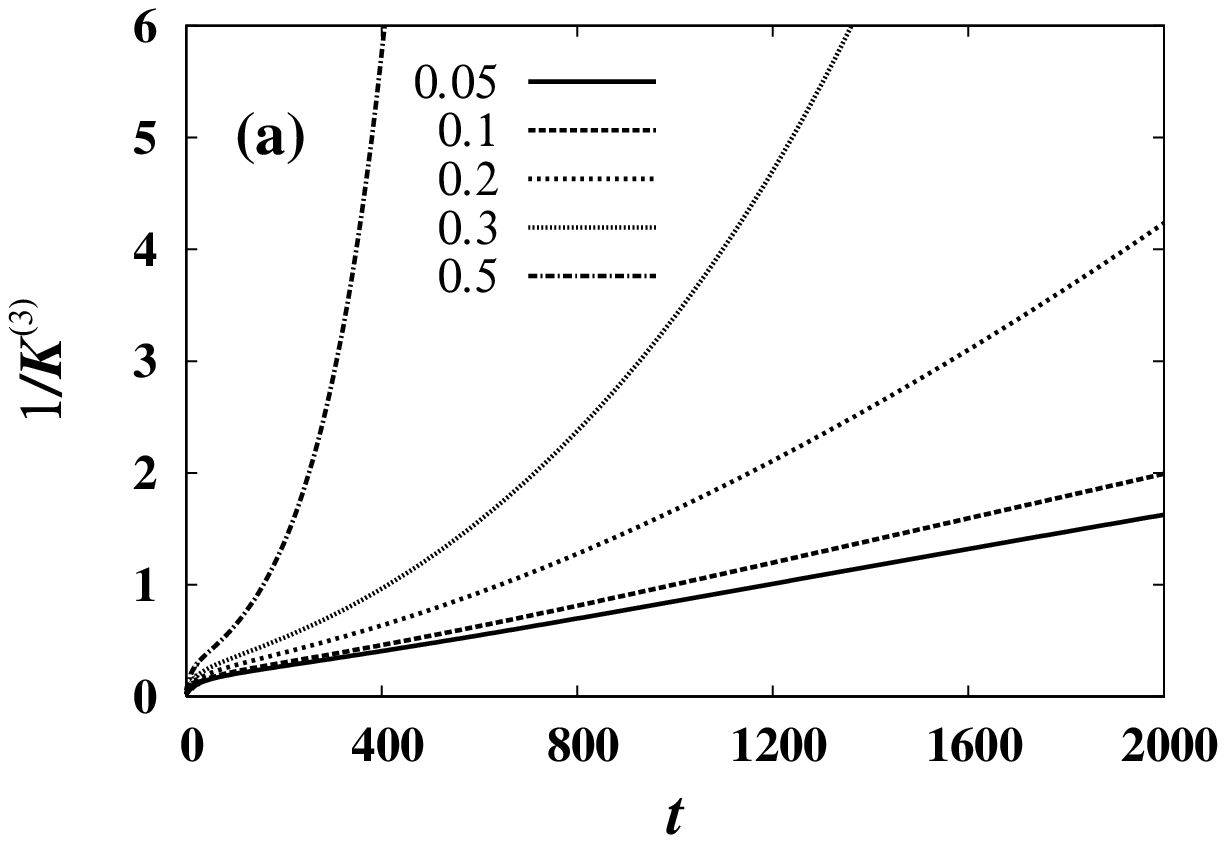}
\includegraphics[width=8.5cm]{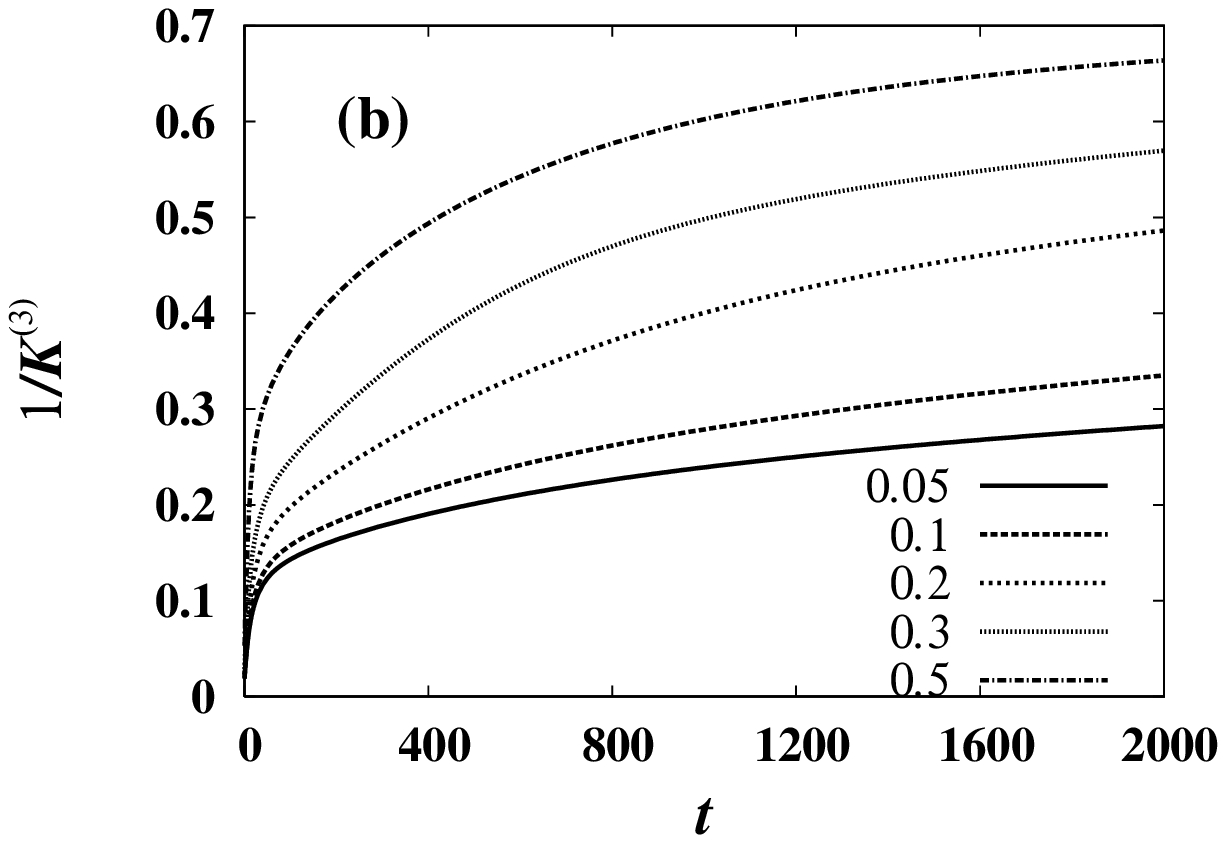}
\caption{Dependences of $1/K^{(3)}$ on time for various 
values of $T/T_{\rm c}$ (numbers in the figure). 
The phonons are supposed to be (a) in thermal equilibrium ($\delta N=0$), 
and (b) in nonequilibrium state ($\delta N\ne 0$). 
}
\label{fig:1}
\end{figure}

The temporal variation of $1/K^{(3)}$ is shown in 
Fig.~\ref{fig:1}. 
If we put $\Delta_0=30$ meV, 
$t=1000$ corresponds to about $22$ ps. 
This is comparable to the range of time taken in the 
experiments~\cite{kaindl}, and we present the 
numerical results for this range hereafter. 
$K^{(3)}$ shows a nonexponential decay for low $T/T_{\rm c}$, 
and it becomes an exponential decay for high $T/T_{\rm c}$ 
as in the experimental result.~\cite{kaindl} 
Here, the nonexponential decay and 
exponential decay indicate 
$1/K^{(3)}\propto 1+\gamma t$ and 
$1/K^{(3)}\propto {\rm e}^{\gamma t}$, respectively. 
At first sight, the former 
seems to be an approximation of the latter 
using ${\rm e}^{\gamma t}\simeq 1+\gamma t$ with $\gamma t\ll 1$, 
but this is not the case. 
For example, at $T/T_{\rm c}=0.05$, 
$1/K^{(3)}$ is fitted to $0.1(1+0.0076t)$ 
($0.0076t\gg 1$ for large $t$). 
Therefore, some qualitative differences 
exist between low and high $T/T_{\rm c}$. 
It is shown below that 
this difference in relaxation dynamics arises 
from the temporal variation of the nonequilibrium 
quasiparticles ($\delta n_{\epsilon}$), and 
this traces back to 
the change of the predominant physical process 
in the collision integral. 
In the case of $\delta N\ne 0$, the relaxation 
becomes slower than that in the case of $\delta N=0$. 
As shown below 
this originates from the existence of $\delta N$, 
which directly affects $K^{(3)}$ 
through the electron-phonon interaction, 
in addition to the slow relaxation of 
nonequilibrium electrons. 
If we vary $g^2N_0$, we obtain 
qualitatively the same results with regard to 
the $t$- and $T$-dependences of $1/K^{(3)}$, but 
there are several quantitative differences. 
In the case of $\delta N=0$ 
$1/K^{(3)}$ increases, 
and the exponential decay of $1/K^{(3)}$ 
is observed at lower $T/T_{\rm c}$ 
with increasing $g^2N_0$. 
This is because 
the equilibration between electrons and phonons 
becomes faster, and $\delta n$ rapidly decreases. 
On the other hand, 
in the case of $\delta N\ne 0$, 
the absolute value of $1/K^{(3)}$ decreases 
with increasing $g^2N_0$. 
In this case, 
the nonequilibrium electrons remain finite, and 
the interaction effect directly enhances $K^{(3)}$. 

\begin{figure}
\includegraphics[width=8.5cm]{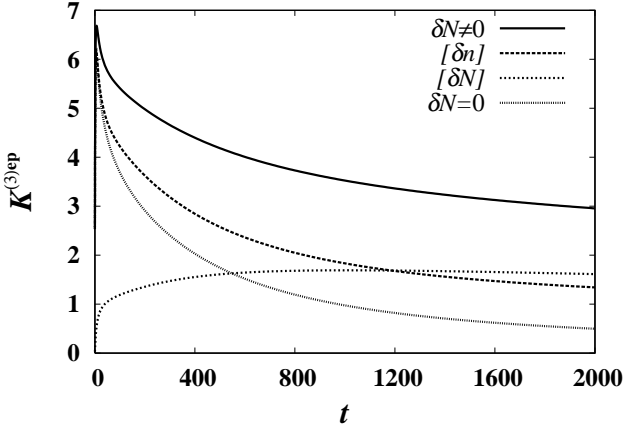}
\caption{Dependences of $K^{(3){\rm el-ph}}$ on time for 
$T/T_{\rm c}=0.1$. 
$\delta N=0$ and $\delta N\ne 0$ indicate 
the results of $K^{(3){\rm el-ph}}$ 
in the cases of $\delta N=0$ and $\delta N\ne 0$, respectively. 
$[\delta n]$ and $[\delta N]$ are components of 
$K^{(3){\rm el-ph}}$ for $\delta N\ne 0$ 
and include only 
$\delta n$ and $\delta N$ 
in $\Sigma^{(a){\rm el-ph}}_{\varphi,\epsilon}$, respectively. 
}
\label{fig:2}
\end{figure}

To clarify the effect of nonequilibrium phonons 
on $K^{(3)}$, we decompose $K^{(3){\rm el-ph}}$ 
to contributions from nonequilibrium electrons 
and phonons 
by taking only the $\delta n$ term or $\delta N$ term 
in $\Sigma^{(a){\rm el-ph}}_{\varphi,\epsilon}$. 
The temporal variations of these terms 
are shown in Fig.~\ref{fig:2}. 
This result indicates that the 
$\delta N$ term has a sufficient contribution 
to $K^{(3)}$ at a large $t$. 
This makes the relaxation slower than 
that in the case of $\delta N=0$. 

\begin{figure}
\includegraphics[width=8.5cm]{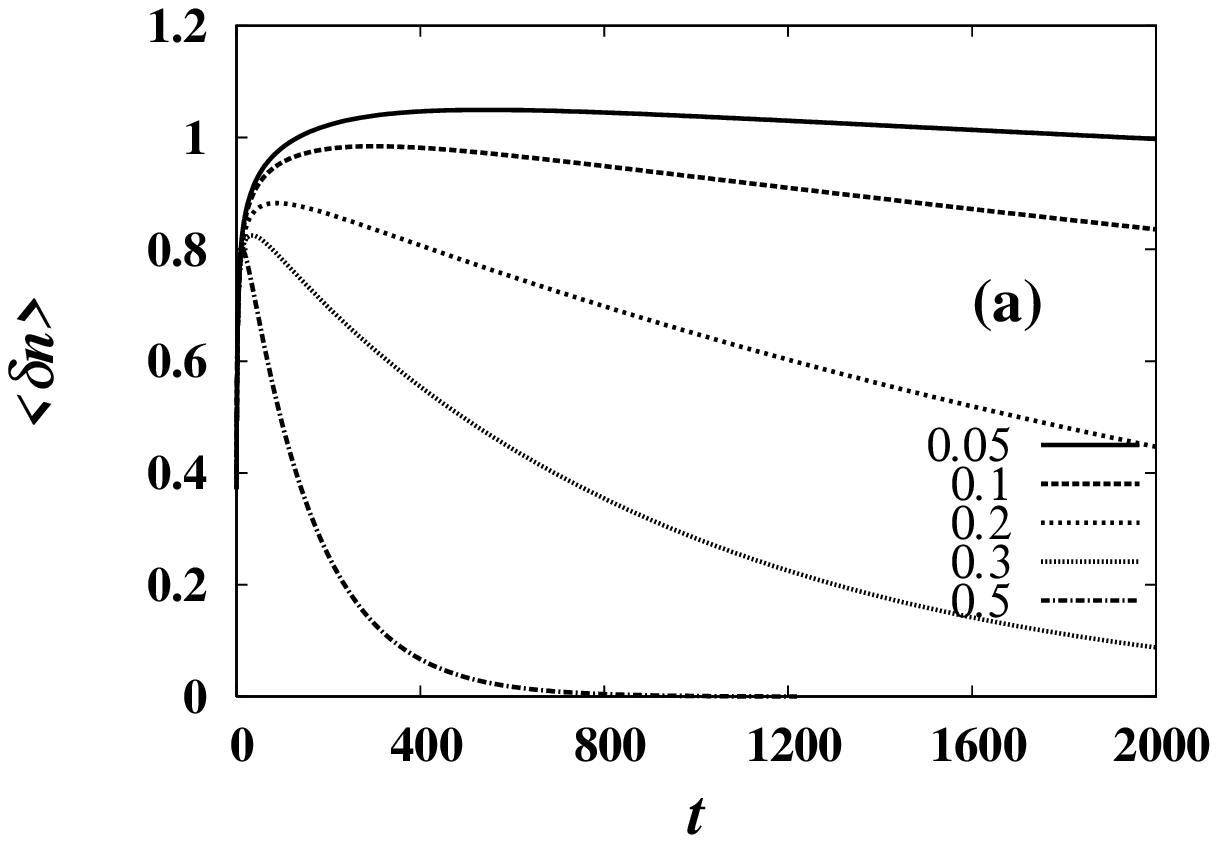}
\includegraphics[width=8.5cm]{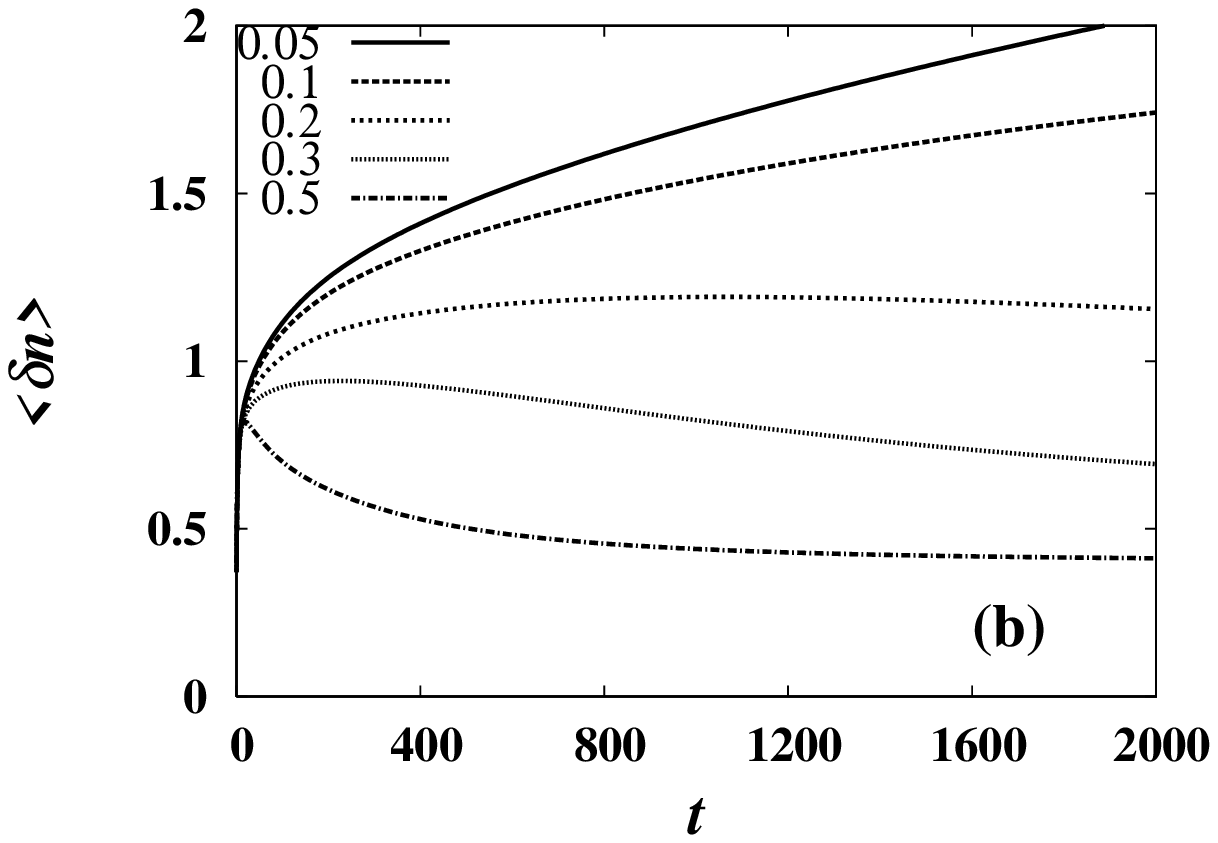}
\caption{Dependences of $\langle\delta n\rangle$ on time 
for various values of $T/T_{\rm c}$ (numbers in the figure). 
The phonons are supposed to be (a) in thermal equilibrium ($\delta N=0$) 
and (b) in a nonequilibrium state ($\delta N\ne 0$). 
}
\label{fig:3}
\end{figure}

Here, we examine the question as to 
whether the nonlinear response 
is proportional to the 
nonequilibrium electron density. 
The temporal variation of the integrated 
nonequilibrium electron density, which is written as 
$\langle\delta n_{\epsilon}(t)\rangle=
\int_0^{\infty}{\rm d}\epsilon\int_{\rm FS}
\frac{2\epsilon}{\sqrt{\epsilon^2-\Delta_{\varphi}^2}}
\delta n_{\epsilon}(t)$, 
is shown in Fig.~\ref{fig:3}. 
In the case of $\delta N=0$, 
$\langle\delta n\rangle$ shows the exponential decay 
for high $T/T_{\rm c}$, as $K^{(3)}$ does. 
For low $T/T_{\rm c}$, $\langle\delta n\rangle$ 
does not show the same $t$-dependence as $K^{(3)}$. 
Therefore, in the region of the nonexponential decay, 
there is no proportionality 
between the nonlinear response 
and nonequilibrium electron density, 
which is different from the assumption 
adopted in previous studies. 
The similarity in relaxation dynamics 
between these two quantities 
is limited to the region of the exponential decay. 
In the case of $\delta N\ne 0$, 
$\langle\delta n\rangle$ does not decrease 
to $0$, but it varies toward the 
nonequilibrium steady state. 
This state is rapidly achieved for high $T/T_{\rm c}$ 
as for $\langle\delta n\rangle$. 
The increase of $\langle\delta n\rangle$ with $t$ 
at low $T/T_{\rm c}$ is understood by considering 
the temporal evolution of $\delta n_{\epsilon}(t)$ 
(shown in Fig. 7) 
and the collision integral $\tilde{I}_{\epsilon}(t)$ 
(in Figs. 6 and 9). 
As discussed there, 
the distribution of nonequilibrium electrons is enhanced 
at low energies by the absorption of phonons. 
(The particle conservation is satisfied 
because $\delta n_{-\epsilon}=-\delta n_{\epsilon}$ 
in our calculation.) 

\begin{figure}
\includegraphics[width=8.5cm]{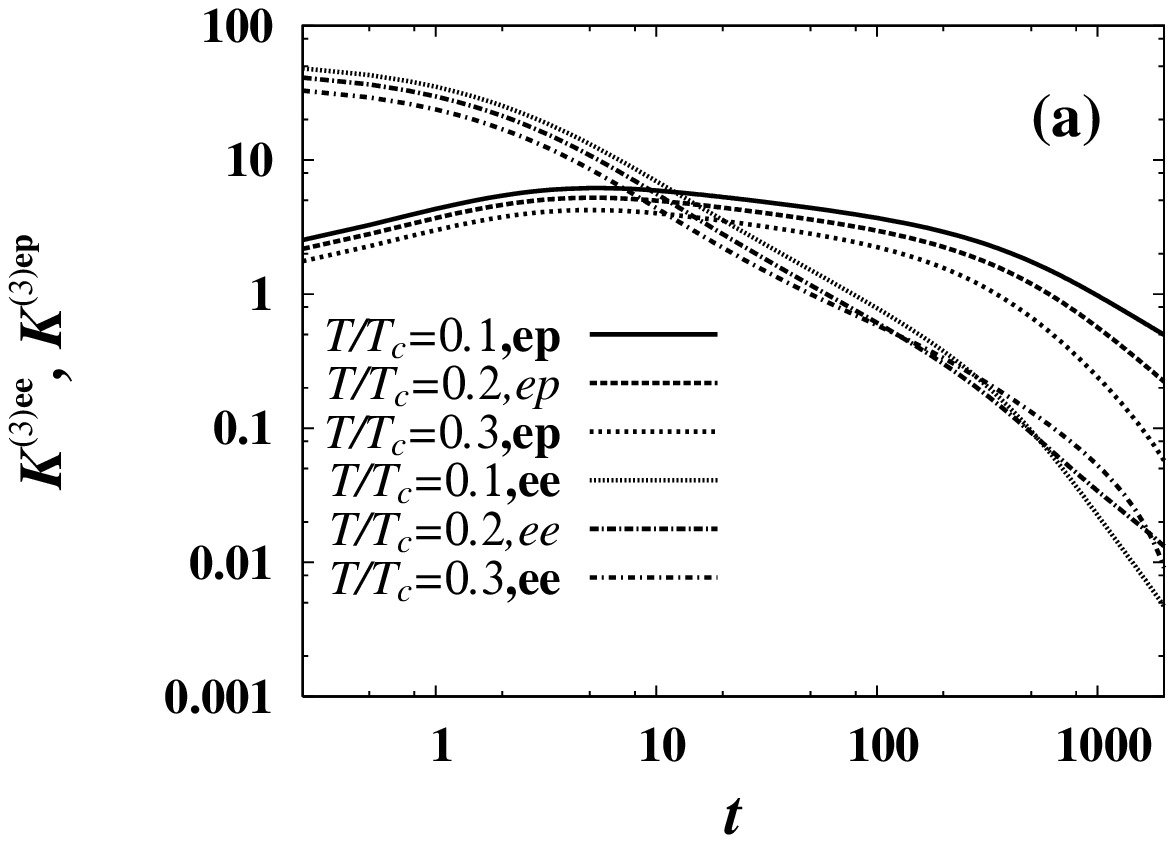}
\includegraphics[width=8.5cm]{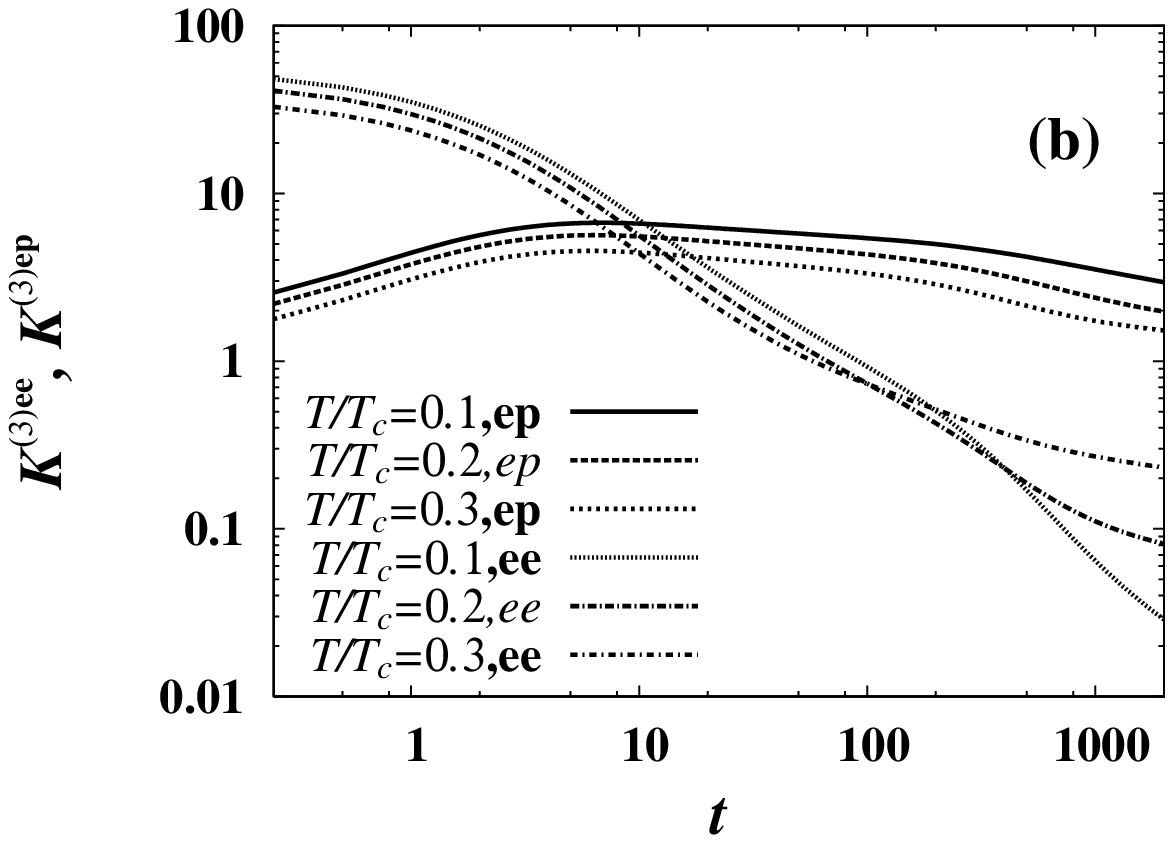}
\caption{Dependences of 
$K^{(3){\rm el-el}}$ ($ee$) and $K^{(3){\rm el-ph}}$ ($ep$) 
on time for various values of $T/T_{\rm c}$. 
The phonons are supposed to be (a) in thermal equilibrium ($\delta N=0$) 
and (b) in a nonequilibrium state ($\delta N\ne 0$). 
}
\label{fig:4}
\end{figure}

Next, we investigate which of the electron-electron and electron-phonon 
interactions predominate in the nonlinear response. 
The temporal variations of $K^{(3){\rm el-ph}}$ and $K^{(3){\rm el-el}}$ 
are shown in Fig.~\ref{fig:4}. 
For small (large) $t$, 
$K^{(3){\rm el-el}}$ ($K^{(3){\rm el-ph}}$) 
is predominant in $K^{(3)}$. 
This change of the predominant term originates 
from the temporal variation of 
the functional form of $\delta n_{\epsilon}(t)$. 
$\delta n_{\epsilon}(t)$ has a broad spectrum 
as a function of $\epsilon$ at small $t$, and 
its spectrum concentrates at low energies 
as time passes, as discussed below. 
The time at which 
$K^{(3){\rm el-el}}$ and $K^{(3){\rm el-ph}}$ 
intersect is weakly dependent on 
whether phonons are in the equilibrium or 
nonequilibrium state, which 
indicates that the effect of nonequilibrium phonons 
is small in this time range. 
Although $K^{(3){\rm el-ph}}$ decreases with increasing $T/T_{\rm c}$, 
$K^{(3){\rm el-el}}$ shows a different behavior at 
large $t$, which also reflects the $\epsilon$-dependence of  
$\delta n_{\epsilon}(t)$, but 
this behavior is not important because 
$K^{(3){\rm el-el}}\ll K^{(3){\rm el-ph}}$ at 
this time scale. 

\begin{figure}
\includegraphics[width=8.5cm]{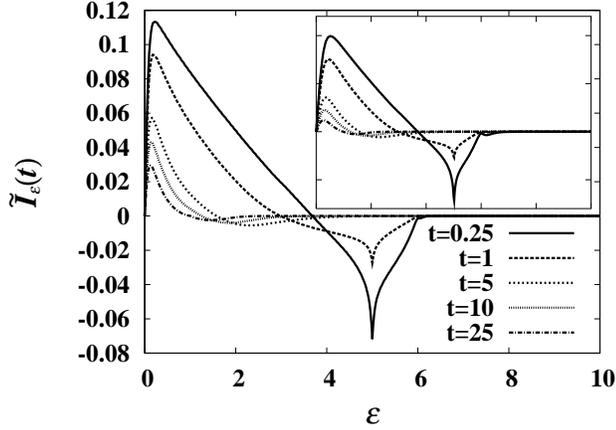}
\caption{Collision integrals $\tilde{I}_{\epsilon}(t)$ 
for various values of $t$. 
$T/T_{\rm c}=0.1$. 
The phonons are supposed to be in thermal equilibrium ($\delta N=0$). 
The results at $T/T_{\rm c}=0.3$ are shown in the inset. 
}
\label{fig:5}
\end{figure}

\begin{figure}
\includegraphics[width=8.5cm]{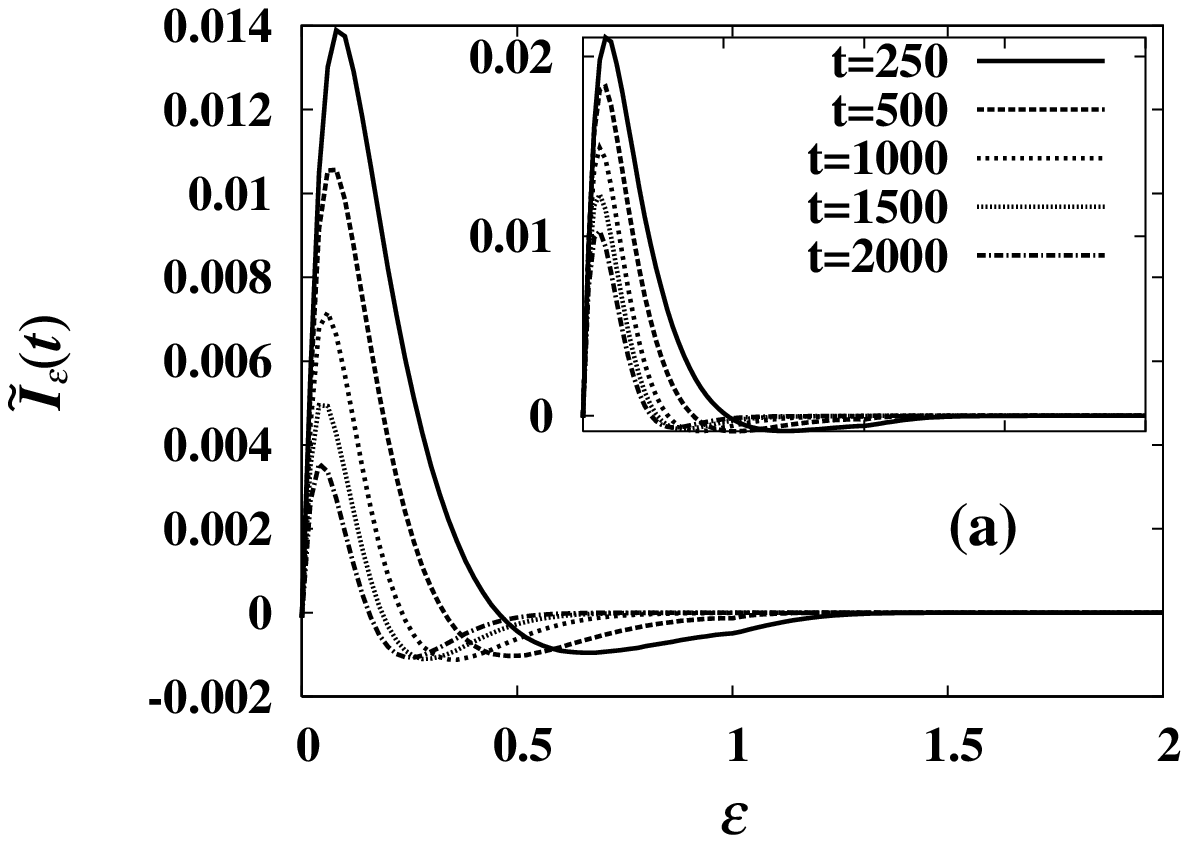}
\includegraphics[width=8.5cm]{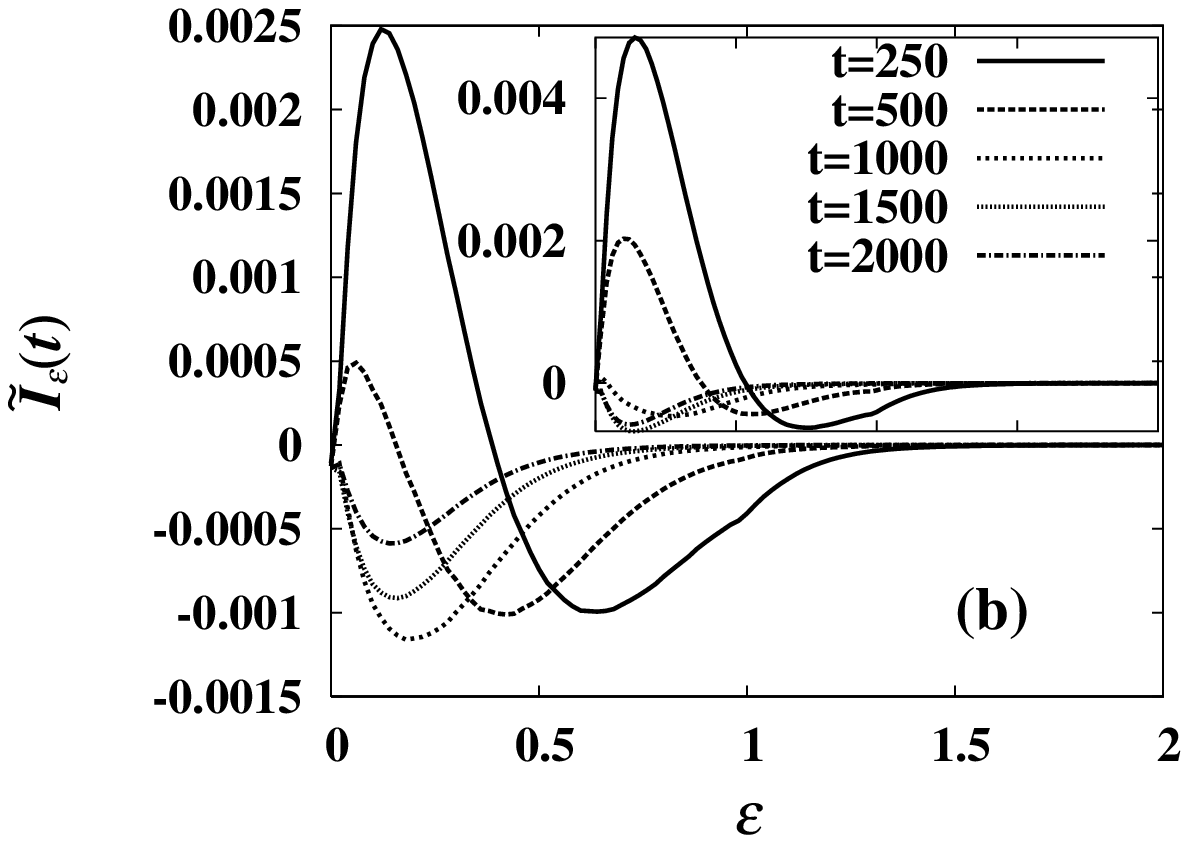}
\caption{Collision integrals $\tilde{I}_{\epsilon}(t)$ 
for various values of $t$. 
(a) $T/T_{\rm c}=0.1$ and (b) $T/T_{\rm c}=0.3$. 
The phonons are supposed to be in thermal equilibrium ($\delta N=0$). 
The results in the case of $\delta N\ne 0$ are 
shown in the inset. 
}
\label{fig:6}
\end{figure}

Hereafter, we consider the microscopic quantities 
that cause the above behavior of the nonlinear response. 
Firstly, we show the temporal variation of 
the collision integral, which is the second 
and third terms on the right side of 
eq. (\ref{eq:nedkineq}) and written as 
$\tilde{I}_{\epsilon}(t)=\tilde{I}^{\rm el-ph}_{\epsilon}(t)
+\tilde{I}^{\rm el-el}_{\epsilon}(t)$ hereafter. 
As indicated above in the cases of 
$K^{(3){\rm el-el}}$ and $K^{(3){\rm el-ph}}$, 
in the collision integral $\tilde{I}_{\epsilon}(t)$, 
the electron-electron interaction predominates 
the relaxation dynamics over 
the electron-phonon interaction 
(roughly $\tilde{I}_{\epsilon}(t)\simeq 
\tilde{I}^{\rm el-el}_{\epsilon}(t)$) at small $t$, and 
it is the other way around 
($\tilde{I}_{\epsilon}(t)\simeq 
\tilde{I}^{\rm el-ph}_{\epsilon}(t)$) at large $t$. 
This holds irrespective of the values of 
$T/T_{\rm c}$ and the existence of $\delta N$. 
We show the collision integral $\tilde{I}_{\epsilon}(t)$ 
(only for $\epsilon \ge 0$ because of 
$\tilde{I}_{-\epsilon}(t)=-\tilde{I}_{\epsilon}(t)$) 
at small $t$ in Fig.~\ref{fig:5} 
and at large $t$ in Fig.~\ref{fig:6}. 
For small $t$, 
there seems to be no qualitative difference in 
the $\epsilon$-dependence of $\tilde{I}_{\epsilon}$ 
between $T/T_{\rm c}=0.1$ and $0.3$. 
The energy range in which $\tilde{I}_{\epsilon}$ 
takes finite values is broad, and 
it takes negative and positive values 
at high and low energies, respectively. 
This brings about a shift in the weight of the 
nonequilibrium distribution 
function from high energy to low energy. 
The results in the case of $\delta N\ne 0$ are omitted here, but 
they are similar to those in the case of $\delta N=0$. 
In contrast to that in the case of small $t$, 
$\epsilon$-dependences 
of $\tilde{I}_{\epsilon}$ are different depending on 
the values of $T/T_{\rm c}$ and $\delta N$ for large $t$. 
For $T/T_{\rm c}=0.3$, $\tilde{I}_{\epsilon}$ becomes 
negative all over the range of $\epsilon>0$. 
This property is the same in the case of $\delta N\ne 0$, 
although its degree is small. 
For $T/T_{\rm c}=0.1$, 
there remains a positive part in 
$\tilde{I}_{\epsilon}$ at low energy. 
The shift in the weight of the nonequilibrium distribution function 
from high energy to low energy occurs, as in the case of small $t$, 
but its energy scale becomes narrower. 
The result for $\delta N\ne 0$ shows 
that the negative part in $\tilde{I}_{\epsilon}$ 
is small and that the decreasing rate of the positive part 
is slower than that of $\delta N=0$. 

\begin{figure}
\includegraphics[width=8.5cm]{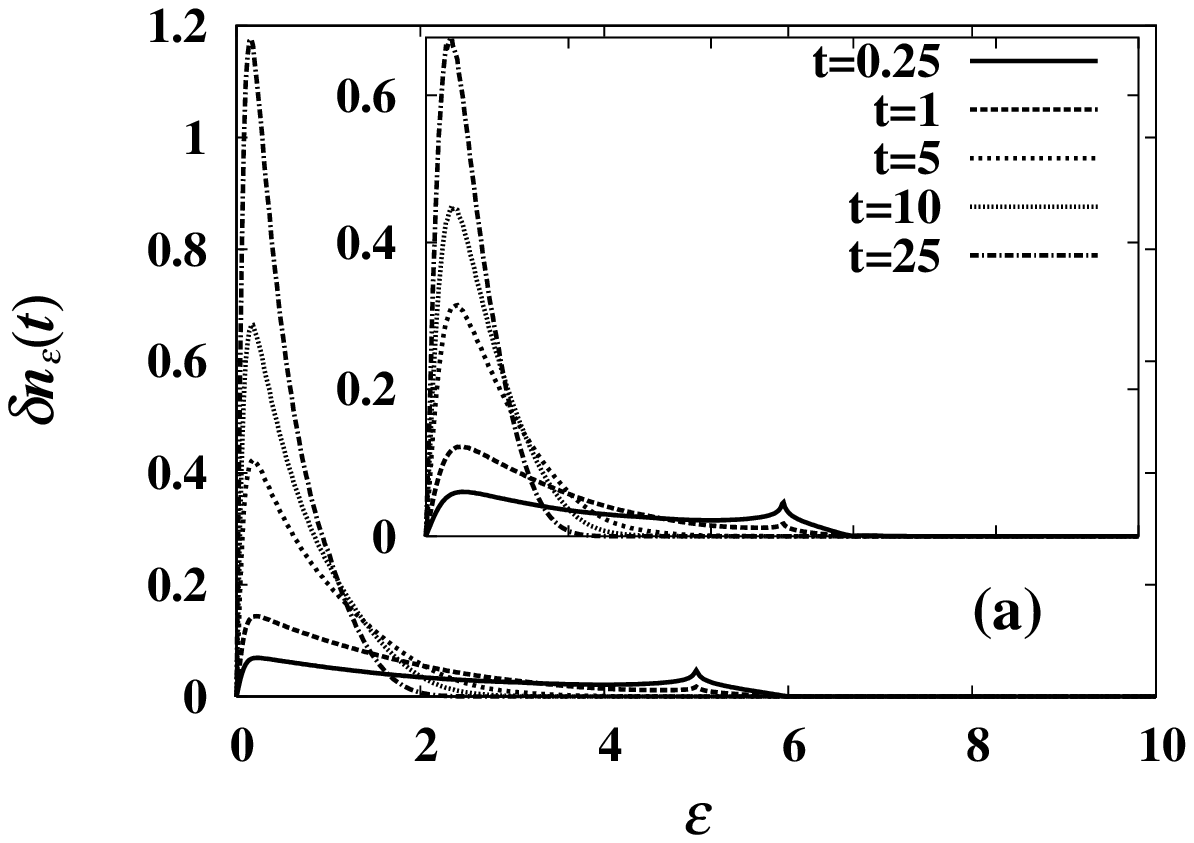}
\includegraphics[width=8.5cm]{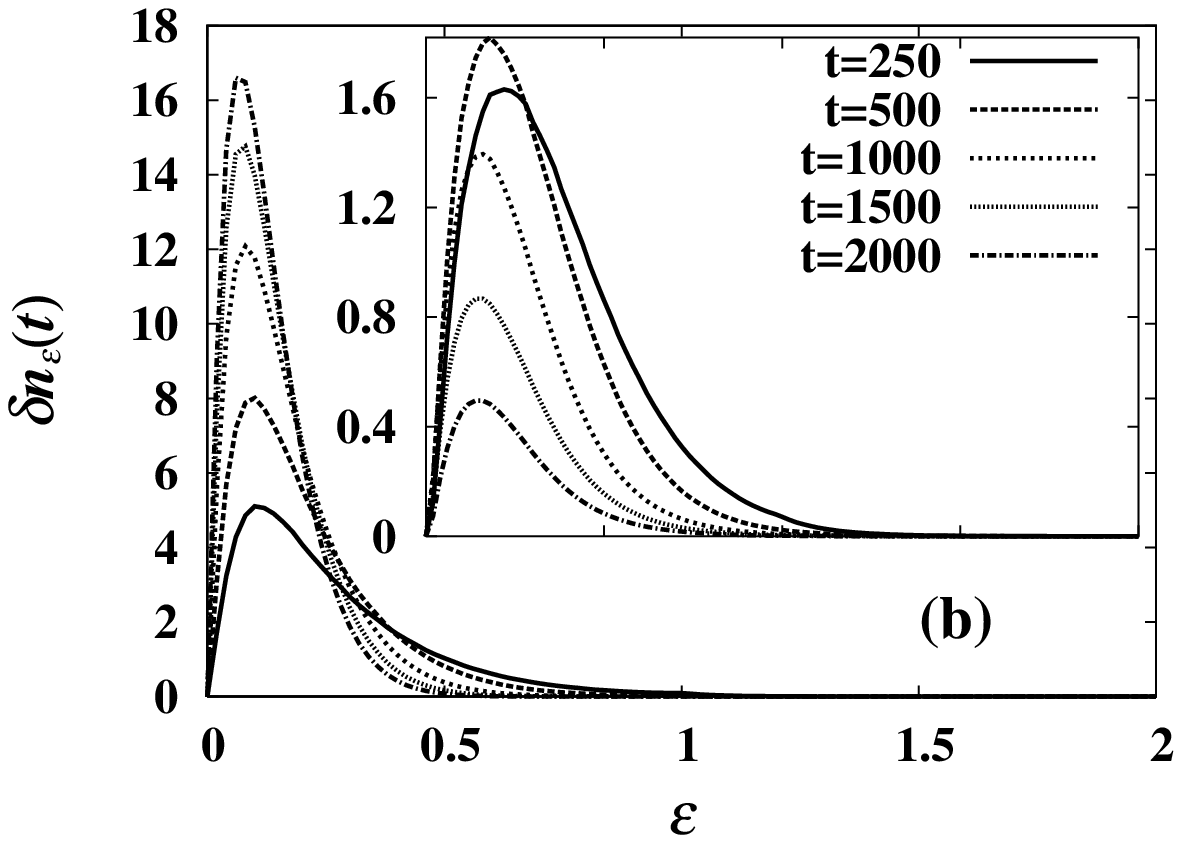}
\caption{Nonequilibrium distribution function 
$\delta n_{\epsilon}(t)$ 
(a) at small values of $t$ and (b) at large values of $t$. 
$T/T_{\rm c}=0.1$. 
The results for $T/T_{\rm c}=0.3$ are shown in the inset. 
The phonons are supposed to be in thermal equilibrium ($\delta N=0$). 
}
\label{fig:7}
\end{figure}

The nonequilibrium distribution functions for 
electrons, $\delta n_{\epsilon}(t)$, 
at various times $t$ 
for $T/T_{\rm c}=0.1$ and $0.3$ in the case of $\delta N=0$ 
are shown in Fig.~\ref{fig:7}. 
(We consider the particle-hole symmetric case: 
$\delta n_{-\epsilon}=-\delta n_{\epsilon}$.) 
At small $t$, immediately after the pump excitation, 
$\delta n_{\epsilon}(t)$ has a broad spectrum. 
The spectrum at high energy rapidly decreases 
with time. 
This is because the damping effect 
is large at high energy owing to the 
electron-electron interaction. 
Then $\delta n_{\epsilon}(t)$ 
takes a functional form, which is similar to that of 
the $T^*$-model~\cite{parker}: 
$\frac{1}{{\rm e}^{\epsilon/T^*}+1}-\frac{1}{{\rm e}^{\epsilon/T}+1}
\simeq \frac{\epsilon(T^*-T)}{[2T{\rm cosh}(\epsilon/2T)]^2}$, 
where $T^*$ is the temperature that characterizes 
nonequilibrium electrons and $T^*>T$. 
For $T/T_{\rm c}=0.1$, 
the shift of the spectrum from high energy to low energy occurs. 
(This indicates 
a nonthermal state, as discussed in ref. 14, 
with taking account of a change in the sign of 
$\partial \delta n_{\epsilon}/\partial t 
\simeq \tilde{I}^{\rm el-ph}_{\epsilon}$ for large $t$.) 
On the other hand, 
$\delta n_{\epsilon}(t)$ starts to decrease all over 
the range of $\epsilon>0$ 
after a certain $t$ for $T/T_{\rm c}=0.3$. 
This behavior is understood by examining 
the $\epsilon$-dependences of the collision integral, as shown above. 
For $\delta N\ne 0$, these tendencies 
are qualitatively the same, which is also indicated in 
the results of the collision integral. 

\begin{figure}
\includegraphics[width=8.5cm]{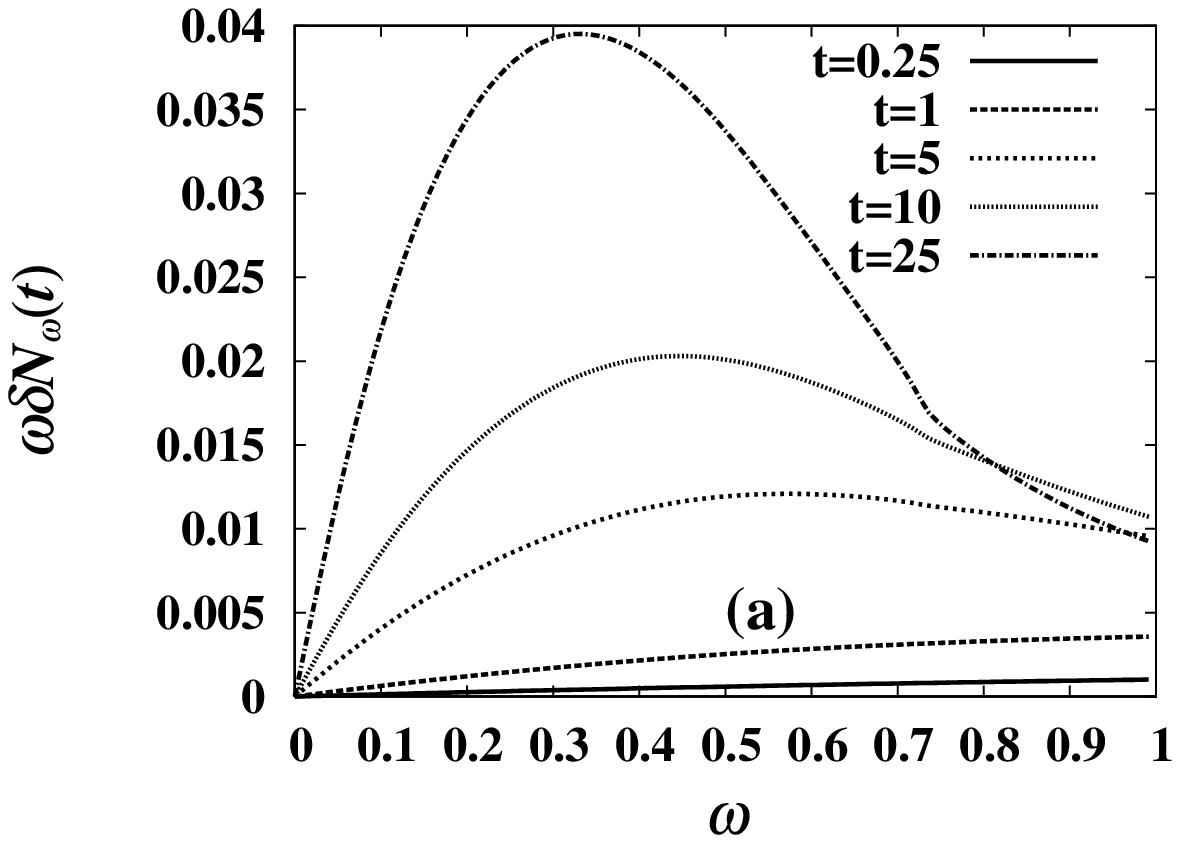}
\includegraphics[width=8.5cm]{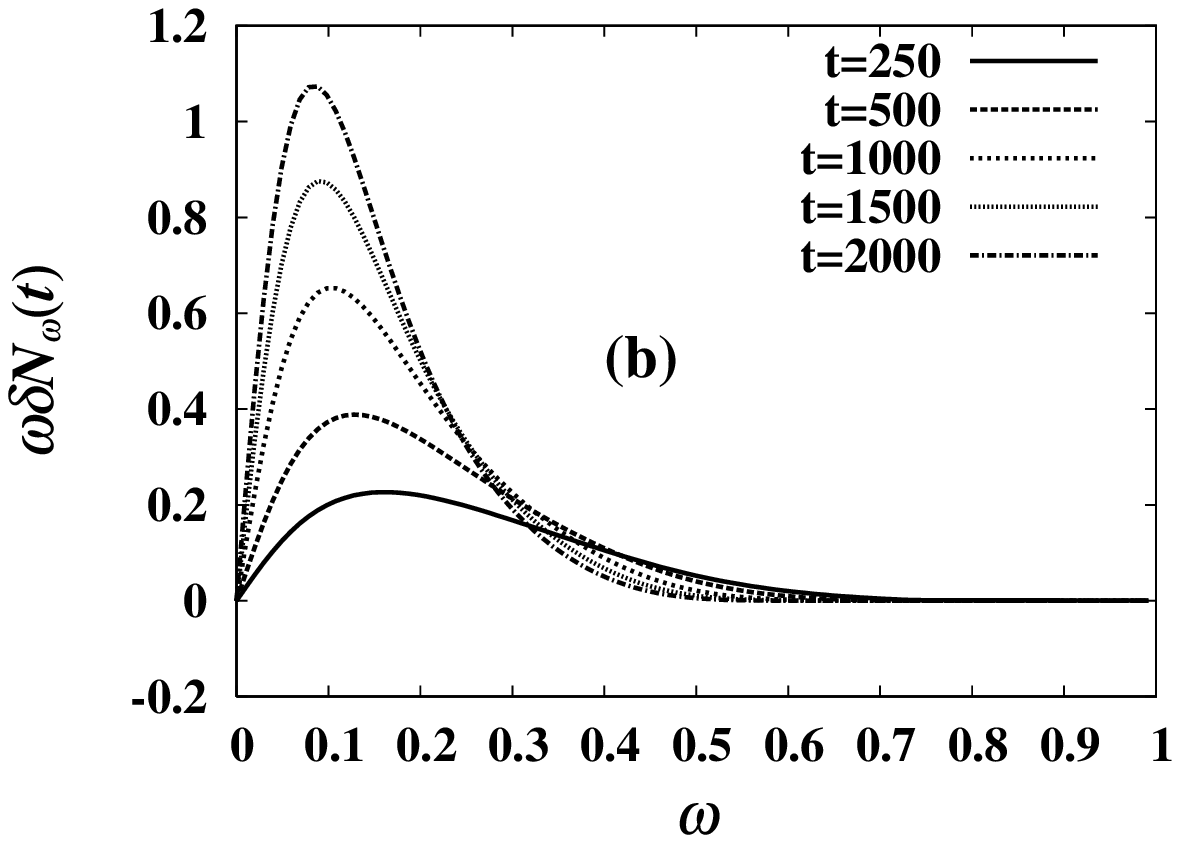}
\caption{$\omega\delta N_{\omega}(t)$ 
(a) at small values of $t$ and (b) at large values of $t$. 
$T/T_{\rm c}=0.1$. 
}
\label{fig:8}
\end{figure}

The energy injected by a pump beam is transferred to 
the phonon system via the electron system. 
In the case of $\delta N=0$, we presume 
that this energy transferred to the phonon system 
dissipates by the damping effect $\gamma_{\rm esc}$ 
and that there is no influence on the electron system 
by $\delta N$. 
On the other hand, 
$\delta N$ is finite if 
the effect of $\gamma_{\rm esc}$ is small. 
The distributions of nonequilibrium phonons 
$\omega\delta N_{\omega}(t)$ 
at various values of $t$ for $T/T_{\rm c}=0.1$ 
are shown in Fig.~\ref{fig:8}. 
The tendency in the changes of the spectrum 
is similar to that observed in nonequilibrium electrons. 
However, the decrease in the spectrum does not occur 
in a phonon system, and nonequilibrium phonons 
accumulate at small $\omega$ with increasing $t$. 
For high $T/T_{\rm c}$, 
$\delta N$ shows a similar but broader spectrum than that 
for low $T/T_{\rm c}$. 

Finally, we investigate which of 
the physical processes causes the nonexponential decay 
of $K^{(3)}$ by increasing $\delta n_{\epsilon}$. 
We rewrite the collision integral by 
the electron-phonon interaction 
$\tilde{I}^{\rm el-ph}_{\epsilon}
:=-\frac{1}{\bar{g}^R_{\epsilon}}
\frac{\pi}{2}N_0g^2\int_{-\phi_D}^{\phi_D}\frac{{\rm d}\phi}{2\pi}
\omega_{\phi}I^{\rm el-ph}_{\phi,\epsilon}
[\delta n_{\epsilon},\delta N_{\phi}]$
to specify the physical processes in this term. 
In the case of $\epsilon > 0$, 
$I^{\rm el-ph}_{\phi,\epsilon}$ is decomposed into 
three terms, 
$I^{\rm el-ph}_{\phi,\epsilon}=
I^a_{\phi,\epsilon}+I^b_{\phi,\epsilon}
+I^c_{\phi,\epsilon}$, in which each term is 
written as follows. 
\[
I^a_{\phi,\epsilon}=
2g^-_{\phi,\epsilon}\delta^{(2)}
[n_{\epsilon}(1-n_{\epsilon-\omega_{\phi}})(1+N_{\omega_{\phi}})
-n_{\epsilon-\omega_{\phi}}(1-n_{\epsilon})N_{\omega_{\phi}}]
\theta(\epsilon-\omega_{\phi}). 
\]
\[
I^b_{\phi,\epsilon}=
-2g^+_{\phi,\epsilon}\delta^{(2)}
[n_{\epsilon+\omega_{\phi}}(1-n_{\epsilon})(1+N_{\omega_{\phi}})
-n_{\epsilon}(1-n_{\epsilon+\omega_{\phi}})N_{\omega_{\phi}}]. 
\]
\[
I^c_{\phi,\epsilon}=
2g^-_{\phi,\epsilon}\delta^{(2)}
[n_{\epsilon}n_{\omega_{\phi}-\epsilon}(1+N_{\omega_{\phi}})
-(1-n_{\omega_{\phi}-\epsilon})(1-n_{\epsilon})N_{\omega_{\phi}}]
\theta(\omega_{\phi}-\epsilon). 
\]
Here, $\delta^{(2)}$ operates $n$ or $N$ 
($(2)$ indicates the order of the external fields), 
and then 
$\delta^{(2)}n_{\epsilon}=\delta n_{\epsilon}(t)$ and  
$\delta^{(2)}N_{\omega_{\phi}}=\delta N_{\omega_{\phi}}(t)$ 
(other $n$ and $N$ are replaced by 
$n^0_{\epsilon}=1/({\rm e}^{\epsilon/T}+1)$ and  
$N^0_{\omega_{\phi}}
=1/({\rm e}^{\omega_{\phi}/T}-1)$, respectively). 
Each term describes 
the emission of phonons ($I^a$), 
the absorption of phonons ($I^b$), and 
the recombination term ($I^c$), respectively. 
(The recombination does not mean that 
quasiparticles recombine into Cooper pairs, 
as the misleading picture in ref. 11 
suggests. 
This term exists in the case of 
the electron-phonon interaction for 
$d$-wave superconductors.) 
We substitute these terms into 
the above collision integral 
$\tilde{I}^{\rm el-ph}_{\epsilon}$, 
and write each term as $\tilde{I}^{a,b,c}_{\epsilon}$. 

\begin{figure}
\includegraphics[width=8.5cm]{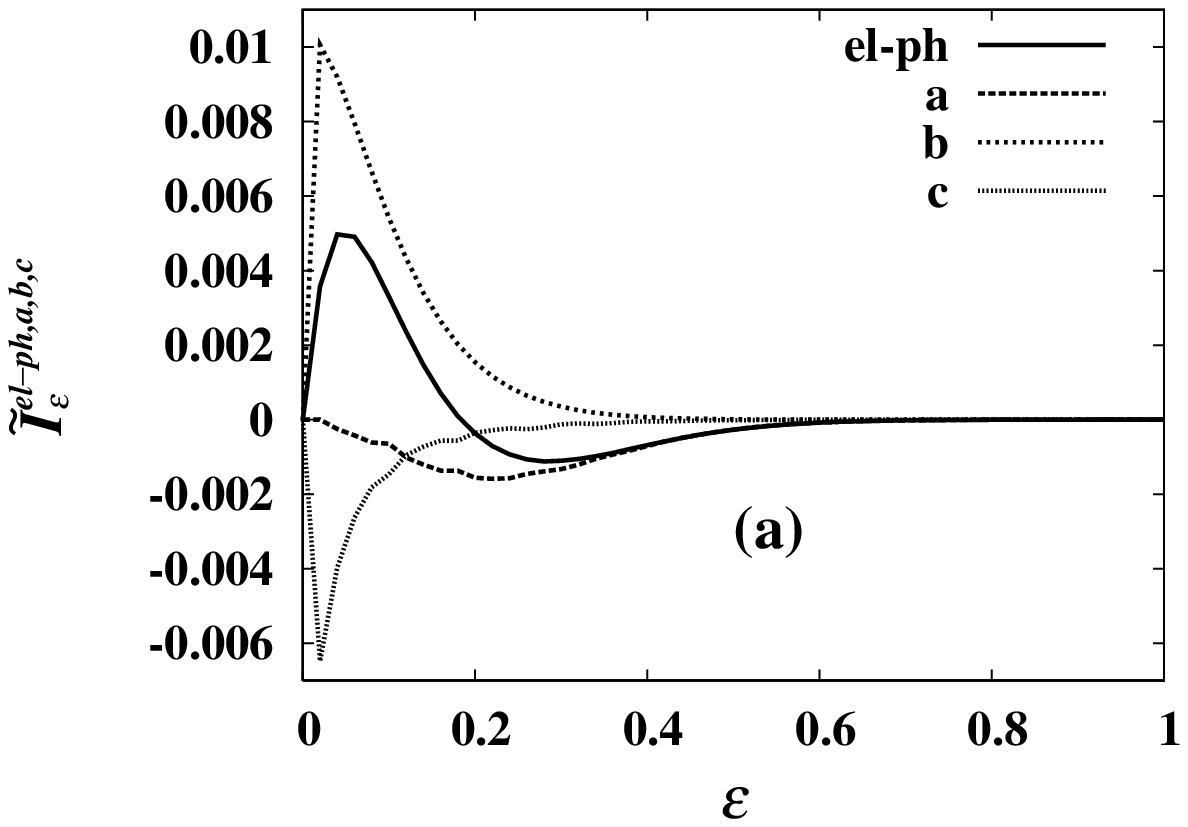}
\includegraphics[width=8.5cm]{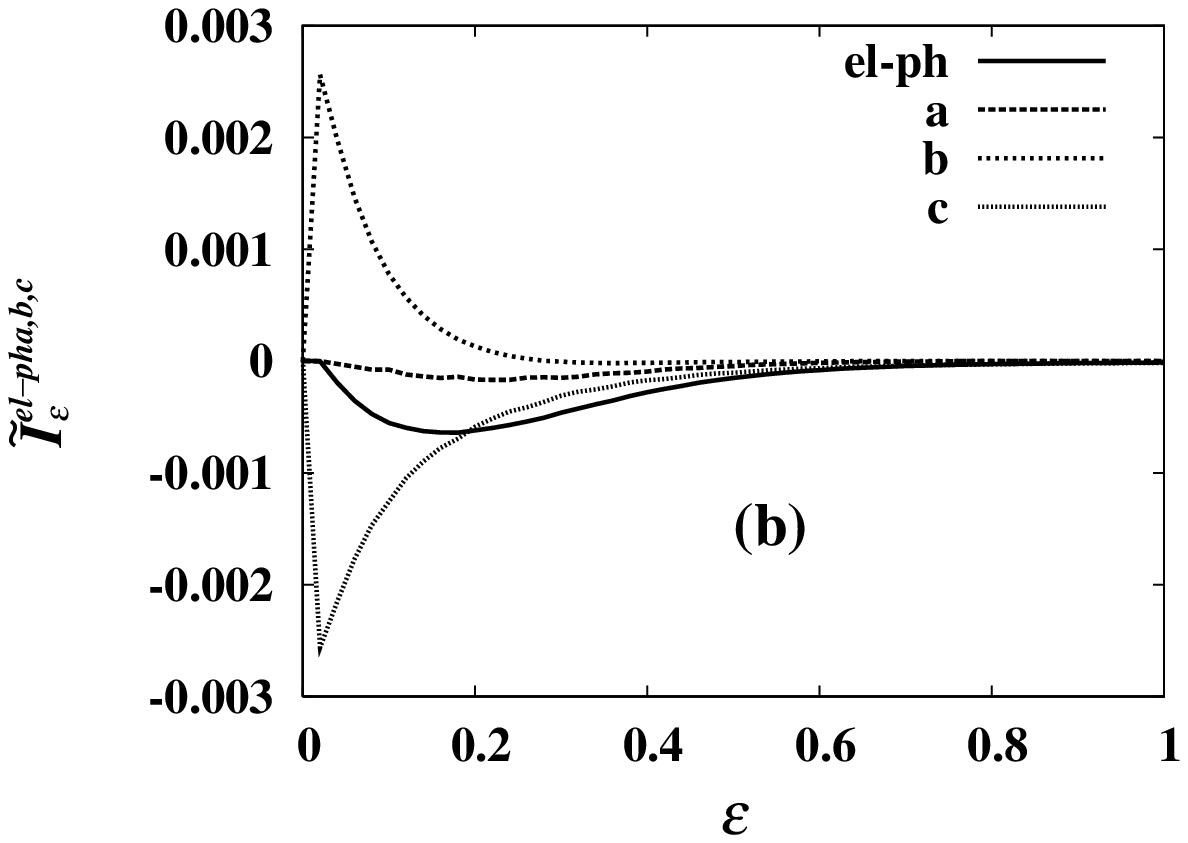}
\caption{Energy dependences of 
$\tilde{I}^{\rm el-ph}_{\epsilon}(t)$ and 
$\tilde{I}^{\rm a,b,c}_{\epsilon}(t)$ 
at $t=1500$ for (a) $T/T_{\rm c}=0.1$ and (b) $T/T_{\rm c}=0.3$.  
The phonons are supposed to be in thermal equilibrium ($\delta N=0$). 
}
\label{fig:9}
\end{figure}

The energy dependences of $\tilde{I}^{\rm el-ph}_{\epsilon}(t)$ and 
$\tilde{I}^{a,b,c}_{\epsilon}(t)$ in the case of $\delta N=0$ 
are shown in Fig.~\ref{fig:9}. 
A scattering term (emission of phonons) 
$\tilde{I}^a_{\epsilon}$ and 
the recombination term $\tilde{I}^c_{\epsilon}$ 
are negative, and 
another scattering term (absorption of phonons)
$\tilde{I}^b_{\epsilon}$ 
is positive. 
At low $T/T_{\rm c}$, 
$\tilde{I}^b_{\epsilon}$ is predominant over 
$\tilde{I}^a_{\epsilon}+\tilde{I}^c_{\epsilon}$, 
and then the collision term 
$\tilde{I}^{\rm el-ph}_{\epsilon}$ 
takes a positive value at low energy. 
Although the absolute values of each term decrease 
with increasing $T/T_{\rm c}$, 
$\tilde{I}^{\rm el-ph}_{\epsilon}$ 
takes a negative value at $T/T_{\rm c}=0.3$ 
because of a rapid decrease in the degree of absorption of 
phonons. 
The negative $\tilde{I}^{\rm el-ph}_{\epsilon}$ for $\epsilon>0$ 
brings about a decrease in 
$\delta n_{\epsilon}(t)$ all over $\epsilon>0$, 
as shown above. 
This causes the exponential decay of $K^{(3)}$. 
On the other hand, the positive 
$\tilde{I}^{\rm el-ph}_{\epsilon}$ at low energy 
increases $\delta n_{\epsilon}(t)$, and then 
reduces the decreasing rate of $K^{(3)}$ as 
compared to that of the exponential decay. 
Therefore, we regard the absorption of phonons 
as the main process for the nonexponential decay 
of $K^{(3)}$. 
This is in contrast to the 
phenomenological interpretation 
based on the RT equation,~\cite{kaindl} 
in which the recombination term 
causes the nonexponential relaxation as the bimolecular decay. 
In the case of $\delta N\ne 0$, 
the energy dependences of $\tilde{I}^{\rm el-ph}_{\epsilon}(t)$ and 
$\tilde{I}^{a,b,c}_{\epsilon}(t)$ are 
qualitatively similar to those of $\delta N=0$. 
However, in this case, the relaxation dynamics 
is slower than that of $\delta N=0$ 
because of the presence of $\delta N$ in $K^{(3)\rm el-ph}$, 
as shown above. 

This predominance of the phonon-absorption term 
at low temperatures is restricted within 
some range of time. 
By performing the calculation further at large $t$, 
it is found that 
the recombination term becomes predominant 
over the absorption of phonons. 
At $T/T_{\rm c}=0.1$, for instance, 
$\tilde{I}^{\rm el-ph}_{\epsilon}(t)$ takes 
negative values in the entire range of $\epsilon>0$ at 
approximately $t\simeq 5000$. 
Then, at about this time, $1/K^{(3)}$ deviates from 
the $t$-linear behavior, and 
ceases to show the nonexponential decay. 

\section{Summary and Discussion}

In this study, we investigated 
the photoinduced change in the superfluid weight 
in the transient state of $d$-wave superconductors. 
We clarified 
how the nonexponential decay occurs. 
At first, we obtained the formula 
that relates the nonequilibrium distribution 
function to the physical response function. 
Although the former quantity is usually 
discussed in theoretical works, 
it is the latter quantity that is necessary 
for a comparison with the experiments. 
In this derivation, we found that 
the vertex correction is predominant 
and that the nonequilibrium distribution function 
is effective only through this interaction term. 
We numerically solved the kinetic equation 
for the nonequilibrium distribution function 
with taking account of the electron-electron and electron-phonon 
interactions and substituted its solution into 
the expression of the nonlinear response function. 

The numerical calculation shows that 
the electron-phonon interaction predominates 
over the electron-electron interaction 
in the long-time behavior. 
In contrast to the previous studies 
based on the phenomenological RT equation, 
the nonexponential decay 
does not originate from the bimolecular recombination. 
Rather, it results from the enhancement of 
the nonequilibrium distribution at low energies, 
which is caused by the absorption of phonons. 
This fact is revealed 
using the nonequilibrium distribution functions 
that couple with each other at different energies 
through the interaction effect, 
and is not known from the RT equation in which the 
quasiparticle density averaged over energy is used. 
This leads to an explanation of 
the nonexponential decay which is 
consistent in terms of the order of 
the external field under the condition 
that the pumping intensity is low. 
(As noted in ref. 17, 
this condition is presumably 
satisfied in the experiment~\cite{kaindl} 
with reference to the excitation fluence 
in ref. 7.) 

Finally, we comment on several problems 
related to but 
beyond the scope of this paper. 
Strictly speaking, 
it requires numerical integration 
with a very small energy scale to discuss 
the long-time behavior of a response function. 
In this paper, however, 
we avoided this by performing 
Fourier transformation at the outset. 
This is achieved at the cost of accuracy in 
the short-time behavior of a response function. 
Therefore, our formulation is not suitable for 
discussing physical quantities such as 
the exact form of the spectrum at this time scale; 
however, the results presented in this paper are not 
considered to be affected by these fine structures 
because of the rapid smoothing by the electron-electron 
interaction. 

There are very few experiments that 
probed low-energy phenomena such as 
the superfluid weight reported in ref. 5. 
Most pump-probe experiments 
have been performed with the use of an optical probe beam. 
The formulation in this paper is restricted to 
the case in which the probe frequency is zero. 
The extension to a finite probe frequency 
is required to discuss optical probe cases, 
but it will not alter the importance of 
the interaction effect if we consider 
the case of the local limit. 

The formulation based on 
the local limit is appropriate for 
cuprate superconductors as discussed in ref. 17. 
Other superconductors are also investigated 
by nonlinear optical spectroscopy, and 
the pump-probe experiments have been performed in 
iron-based superconductors 
(for example, ref. 24.) 
Infrared spectroscopy indicates that 
the nonlocal limit is realized 
in superconductors of this kind,~\cite{li} 
and in this case 
the Mattis-Bardeen formula~\cite{mattis} is valid. 
It is possible that 
the nonlocal limit gives 
a different result regarding the nonlinear response 
from the local limit in which 
the interaction effect is essential 
for the optical response. 

\section*{Acknowledgement}
The numerical calculations were carried out on SX8 
at YITP in Kyoto University.


\begin{thebibliography}{9}

\bibitem{averitt} R. D. Averitt, G. Rodriguez, A. I. Lobad, 
J. L. W. Siders, S. A. Trugman, and A. J. Taylor: 
Phys. Rev. B {\bf 63} (2001) 140502(R). 

\bibitem{segre} G. P. Segre, N. Gedik, J. Orenstein, D. A. Bonn, 
Ruixing Liang, and W. N. Hardy: 
Phys. Rev. Lett. {\bf 88} (2002) 137001. 

\bibitem{dvorsek} D. Dvorsek, V. V. Kabanov, J. Demsar, 
S. M. Kazakov, J. Karpinski, and D. Mihailovic: 
Phys. Rev. B {\bf 66} (2002) 020510. 

\bibitem{gedik04} N. Gedik, P. Blake, R. C. Spitzer, 
J. Orenstein, R. Liang, D. A. Bonn, and W. N. Hardy: 
Phys. Rev. B {\bf 70} (2004) 014504. 

\bibitem{kaindl} R. A. Kaindl, M. A. Carnahan, D. S. Chemla, 
S. Oh, and J. N. Eckstein: 
Phys. Rev. B {\bf 72} (2005) 060510(R). 

\bibitem{perfetti} L. Perfetti, P. A. Loukakos, M. Lisowski, 
U. Bovensiepen, H. Eisaki, and M. Wolf: 
Phys. Rev. Lett. {\bf 99} (2007) 197001. 

\bibitem{kusar} P. Kusar, V. V. Kabanov, J. Demsar, T. Mertelj, 
S. Sugai, and D. Mihailovic: 
Phys. Rev. Lett. {\bf 101} (2008) 227001. 

\bibitem{giannetti} C. Giannetti, G. Coslovich, F. Cilento, G. Ferrini, 
H. Eisaki, N. Kaneko, M. Greven, and F. Parmigiani: 
Phys. Rev. B {\bf 79} (2009) 224502. 

\bibitem{kabanov} V. V. Kabanov, J. Demsar, B. Podobnik, 
and D. Mihailovic: 
Phys. Rev. B {\bf 59} (1999) 1497. 

\bibitem{rothwarf} A. Rothwarf and B. N. Taylor: 
Phys. Rev. Lett. {\bf 19} (1967) 27. 

\bibitem{chang77} J.-J. Chang and D. J. Scalapino: 
Phys. Rev. B {\bf 15} (1977) 2651. 

\bibitem{ovchinnikov} Y. N. Ovchinnikov and V. Z. Kresin: 
Phys. Rev. B {\bf 58} (1998) 12416. 

\bibitem{kabanov05} V. V. Kabanov, J. Demsar, 
and D. Mihailovic: 
Phys. Rev. Lett. {\bf 95} (2005) 147002. 

\bibitem{ahn} K. H. Ahn, M. J. Graf, S. A. Trugman, J. Demsar, 
R. D. Averitt, J. L. Sarrao, and A. J. Taylor: 
Phys. Rev. B {\bf 69} (2004) 045114. 

\bibitem{groeneveld} R. H. M. Groeneveld, R. Sprik, 
and A. Lagendijk: 
Phys. Rev. B {\bf 51} (1995) 11433. 

\bibitem{kabanov08} V. V. Kabanov and A. S. Alexandrov: 
Phys. Rev. B {\bf 78} (2008) 174514. 

\bibitem{jujo10} T. Jujo: 
J. Phys. Soc. Jpn. {\bf 79} (2010) 034705. 

\bibitem{jujo07} T. Jujo: 
J. Phys. Soc. Jpn. {\bf 76} (2007) 073703. 

\bibitem{note1} 
If the self-energy is independent of energy and 
the vertex correction does not exist, 
there is no contribution from $\tilde{g}^{(a)(3)}_{\epsilon,\epsilon}$ 
to $K^{(3)}$, and then 
only $\tanh\tfrac{\epsilon}{2T}
(g^{R(3)}_{\epsilon,\epsilon}-g^{A(3)}_{\epsilon,\epsilon})$ 
remains.~\cite{jujo10}  
However, this term does not show decay 
in relaxation dynamics. 
The fact that decay occurs in the experiments 
indicates the existence of the energy-dependence on 
the self-energy and vertex correction. 
This also shows a predominance of 
$\tilde{g}^{(a)(3)}_{\epsilon,\epsilon}$ over another term, 
which is consistent with the calculation. 

\bibitem{eliashberg} G. M. \'Eliashberg: 
Zh. Eksp. Teor. Fiz. {\bf 61} (1971) 1254. 
(Sov. Phys. JETP {\bf 34} (1972) 668.) 

\bibitem{gulian} A. M. Gulian and G. F. Zharkov: 
{\it Nonequilibrium Electrons and Phonons in 
Superconductors}
(Kluwer Academic/Plenum Publishers, New York, 1999). 

\bibitem{note2} 
The value of $\Omega_0$ is small 
compared with the experimental value (about 50), but 
the quasiparticle density excited at high energy 
rapidly decreases because of the electron-electron interaction, 
and then nonequilibrium electrons accumulate at low energy 
as the numerical calculation shows. 
Therefore, this value of $\Omega_0$ does not affect 
the results below. 

\bibitem{parker} W. H. Parker: 
Phys. Rev. B {\bf 12} (1975) 3667. 

\bibitem{mertelj} T. Mertelj, V. V. Kabanov, C. Gadermaier, 
N. D. Zhigadlo, S. Katrych, J. Karpinski, and D. Mihailovic: 
Phys. Rev. Lett. {\bf 102} (2009) 117002. 

\bibitem{li} G. Li, W. Z. Hu, J. Dong, Z. Li, P. Zhang, 
G. F. Chen, J. L. Luo, and N. L. Wang: 
Phys. Rev. Lett. {\bf 101} (2008) 107004. 

\bibitem{mattis} D. C. Mattis and J. Bardeen: 
Phys. Rev. {\bf 111} (1958) 412. 

\end{thebibliography}
\end{document}